\documentclass[12pt]{iopart}

\usepackage{iopams}
\usepackage{graphicx}

\begin{document}

\title{The quantum Rabi model: solution and dynamics}

\author{Qiongtao Xie$^{1}$, Honghua Zhong$^{2,3,4}$, Murray T. Batchelor$^{5,6}$ and Chaohong Lee$^{2,3,7}$}
\address{$^{1}$College  of Physics and Electronic Engineering, Hainan Normal University, Haikou 571158, People's Republic of China}
\address{$^{2}$TianQin Research Center \& School of Physics and Astronomy, Sun Yat-Sen University (Zhuhai Campus), Zhuhai 519082, People's Republic of China}
\address{$^{3}$State Key Laboratory of Optoelectronic Materials and Technologies, Sun Yat-Sen University (Guangzhou Campus), Guangzhou 510275, People's Republic of China}
\address{$^{4}$Department of Physics, Jishou University, Jishou 416000, People's Republic of China}
\address{$^{5}$Centre for Modern Physics, Chongqing University, Chongqing 400044, People's Republic of China}
\address{$^{6}$Mathematical Sciences Institute and Department of Theoretical Physics, Research School of Physics and Engineering, Australian National University, Canberra ACT 0200, Australia}

\address{$^{7}$Author to whom correspondence should be addressed.}

\ead{lichaoh2@mail.sysu.edu.cn}


\begin{abstract}
This article presents a review of recent developments on various aspects of the quantum Rabi model.
Particular emphasis is given on the exact analytic solution obtained in terms of confluent Heun functions.
The analytic solutions for various generalisations of the quantum Rabi model are also discussed.
Results are also reviewed on the level statistics and the dynamics of the quantum Rabi model.
The article concludes with an introductory overview of several experimental realisations of the quantum Rabi model.
An outlook towards future developments is also given.
\end{abstract}



\section{Introduction}

80 years ago Rabi~\cite{Rabi} introduced a model to discuss the effect of a rapidly varying weak magnetic field
on an oriented atom possessing nuclear spin.
The simplest case corresponds to a two-state system.
The atom is treated as quantised, and the field is treated as a classically rotating field.
The effect of a non-rotating, alternating field was later discussed by Bloch and Siegert \cite{BS}, who
found a shift in the position of the resonance -- now called the Bloch-Siegert shift.
This shift has been observed in experiments with a driven superconducting qubit \cite{TSSTMH}.

Jaynes and Cummings~\cite{JC} introduced a similar quantum model in 1963 describing a two-level atom interacting
with a quantised mode of an optical cavity.
Their initial goal was to study the relationship between the
quantum theory of radiation and the corresponding semi-classical theory.
Despite its simplicity, the quantum Rabi model was not regarded as exactly solvable.
To solve the model, a rotating wave approximation (RWA) was taken.
In this approximation, known as the Jaynes-Cummings (JC) model, the counter-rotating term (CRT) is neglected,
which turned out to be
a valid approximation for the near resonance and weak coupling parameter regions of relevance to many experiments.
The JC model is readily solved, including the dynamics,
and has been very successfully applied to understand a range of experimental phenomena,
such as vacuum Rabi mode splitting~\cite{Thompson}
and quantum Rabi oscillation~\cite{Brune}.\footnote{For an early review of JC physics,
the reader is referred to \cite{SN} (see also \cite{Nobel2012a,Nobel2012b}).}

The various coupling regimes of the quantum Rabi model can be defined in terms of the qubit frequency $2 \Delta$,
the mode frequency $\omega$ and the coupling $g$ between the two systems.
These regimes are~\cite{Pedernales}\footnote{One further regime is defined by the limit $\omega=0$,
which is the relativistic Dirac regime~\cite{Pedernales}.}
\begin{enumerate}
\item decoupling regime: $2\Delta \ll g \ll \omega$,
\item JC regime: $g \ll \omega, 2\Delta$ and $|\omega - 2\Delta| \ll |\omega + 2\Delta|$,
\item anti JC regime: $g \ll \omega, 2\Delta$ and $|\omega - 2\Delta| \gg |\omega + 2\Delta|$,
\item intermediate regime: $2\Delta \sim g \ll \omega$,
\item two-fold dispersive regime: $g < \omega, 2\Delta, |\omega - 2\Delta|, |\omega + 2\Delta|$,
\item ultrastrong coupling regime: $0.1 < g < \omega$,
\item deep strong coupling regime: $g > \omega$.
\end{enumerate}

In important experimental developments for engineered quantum systems,
all relevant system parameters are tunable, allowing new quantum  regimes to be reached.
These systems include superconducting qubits coupled to
microwave waveguide resonators \cite{Abdumalikov,Fink, Niemczyk,Bourassa},
LC resonators \cite{Johansson,Mooij,Fedorov} and mechanical resonators \cite{LaHaye,OConnell,Pirkkalainen}.
In particular, the ultrastrong coupling regime can be realised.
In addition, in femtosecond-laser-written waveguide superlattices, a classical simulator of the
quantum Rabi model in the deep strong coupling regime has been realised~\cite{Crespi}.
Moreover, results have been reported for a superconducting qubit-oscillator circuit
in the ultra strong coupling regime and beyond~\cite{Forn-Diaz,Yoshihara}.
In such regimes, the usual RWA is no longer valid and the CRT cannot be neglected.
The direct evidence of the breakdown of the JC model has been reported~\cite{Niemczyk}.
Various methods have been proposed to tackle the strong coupling regimes, including
what has come to be known as
a generalised RWA~\cite{FKU,I2007,Albert2011,YZLCJ,ZhangCZ2013,ZhangChen2015,ZCHC}
for obtaining successive approximations for the eigenspectrum of the quantum Rabi model.
As a result, several interesting phenomena due to the CRT
have been predicted (see also~\cite{Wolfa,Wolfb, ZCZ2013,Casanova,He,Garziano2015,Garziano2016}).

In another development, it was found by Braak in 2011 that the quantum Rabi model is exactly solvable.
Braak presented an analytic solution for the quantum Rabi model in the Bargmann-Fock space of analytic functions,
deriving conditions for determining the energy spectrum~\cite{Braak,Braakb,Braak_review}.
Subsequently this condition was reproduced via Bogoliubov transformation~\cite{CWHLW2012}.
It was further found that the analytic solution for the quantum Rabi model can be given in
terms of confluent Heun functions \cite{Zhong1,Maciejewskia}, with
the well-known Judd isolated exact solutions \cite{Judd} appearing naturally as
truncations of the infinite series defining the confluent Heun functions.
Braak's analytic solution of the quantum Rabi model heralded an ongoing wave of
solutions for the full eigenspectrum of various known generalisations of the quantum Rabi model.
The emphasis of this article is on reviewing these developments,
with particular attention given to the analytic solutions.

The review is set out as follows.
In Section 2 we discuss the eigenvalue problem and analytic solutions obtained for the quantum Rabi model, also
touching on the issue of integrability.
Section 3 is devoted to the energy spectrum and dynamics of the quantum Rabi model,
including both the regular and exceptional parts of the eigenspectrum.
Level statistics and dynamics of the quantum Rabi model are also discussed in this section.
In Section 4 we discuss the analytic solutions obtained for the asymmetric quantum Rabi model, the
anisotropic quantum Rabi model and the two-photon quantum Rabi model.
Section 5 is devoted to an overview of a selection of experimental realisations of the quantum Rabi model.
The different experimental platforms covered are a single atom in a cavity,
superconducting circuits and hybrid mechanical systems.
Concluding remarks, with an outlook to future developments, are given in Section 6.

\section{Eigenvalue problem and analytic solutions}

The quantum Rabi model is described by the hamiltonian $(\hbar=1)$
\begin{equation}
H_{\mathrm{R}}=\Delta
\sigma_z+\omega a^{\dagger}a+g\sigma_x(a^{\dagger}+a),\label{Rabiha1}
\end{equation}
where $a$ and $a^{\dagger}$ are the destruction and creation operators
for a single bosonic mode of frequency $\omega$, $\sigma_{x}$ and $\sigma_{z}$ are
Pauli matrices for a two-level system with level splitting $2\Delta$, and
$g$ denotes the interaction between the two systems.
In terms of the spin raising and lowering operators $\sigma^{\pm}=\frac12 (\sigma_x\pm {\mathrm i} \, \sigma_y)$
the interaction term can be written as the sum of two terms:
$g\sigma_x(a^{\dagger}+a)=g( \sigma^-a^\dagger+ \sigma^+a)+g( \sigma^+a^\dagger+ \sigma^-a)$,
where $g(a^\dagger \sigma^-+a \sigma^+)$ is the rotating term and
$g(a^\dagger \sigma^++a \sigma^-)$ is the CRT.
Jaynes and Cummings  \cite{JC} proposed the RWA where the CRT is neglected.
The resulting simplified model, with hamiltonian
\begin{equation}
H_{\mathrm{JC}}=\Delta
\sigma_z+\omega a^{\dagger}a+g( \sigma^-a^\dagger+ \sigma^+a),\label{JCha1}
\end{equation}
is known as the Jaynes-Cummings (JC) model. This model has the additional conserved operator
\begin{equation}
N=a^\dagger a+\sigma^+\sigma^-,
\end{equation}
which can be readily seen to commute with $H_{\mathrm{JC}}$.
As a result, the state space of the JC Hamiltonian decomposes into an infinite sum of two-dimensional invariant subspaces.
The operator $N$ generates a continuous $U(1)$ symmetry of the JC model.
In the quantum Rabi model, the  $U(1)$ symmetry is broken by the CRT.
Nevertheless, the quantum Rabi model does have a $Z_2$ symmetry, which has been used in the
derivation of conditions determining the energy spectrum \cite{Braak,Braakb}.
This symmetry is further discussed in~\cite{Albert2012,Gardas,Wak}.
For the quantum Rabi model, the parity operator
\begin{equation}
\Pi=-\sigma_z(-1)^{a^\dagger a}, \label{parity}
\end{equation}
is the conserved quantity, with $[H_R,\Pi]=0$ and eigenvalues $p=\pm 1$.

The eigenstate $\left|\psi\right\rangle$ of the quantum Rabi hamiltonian can be expressed as two-component wave functions
\begin{equation}
\left|\psi\right\rangle=\left(
\begin{array}{c}
\psi_1 \\
\psi_2
\end{array}%
\right).
\end{equation}
From the Schr\"{o}dinger equation $H_{\mathrm{R}}|\psi\rangle=E|\psi\rangle$, it follows
\begin{eqnarray}
 a^{\dagger}a \, \psi_1+g(a^{\dagger}+a)\psi_2+\Delta \psi_1 &=&E\psi_1,\\
  a^{\dagger}a \, \psi_2-g(a^{\dagger}+a)\psi_1-\Delta \psi_2 &=&E\psi_2.
\end{eqnarray}
Here for brevity, we have set $\omega=1$.
Introducing linear combinations of $\psi_{1}$ and $\psi_{2}$,
$\phi_1=\psi_1+\psi_2$ and $\phi_2=\psi_1-\psi_2$, gives
\begin{eqnarray}
 a^{\dagger}a \, \phi_1+g(a^{\dagger}+a)\phi_1+\Delta \phi_2 &=&E\phi_1,\\
  a^{\dagger}a \, \phi_2+g(a^{\dagger}+a)\phi_2+\Delta \phi_1 &=&E\phi_2.
\end{eqnarray}
This can be written in  the matrix form
\begin{equation}
H_{\mathrm{R}}'\left(
\begin{array}{c}
\phi_1 \\
\phi_2
\end{array}%
\right)=E\left(
\begin{array}{c}
\phi_1 \\
\phi_2
\end{array}%
\right)\label{eigenvalue}
\end{equation}
with
\begin{equation}
H_{\mathrm{R}}'=\left(
\begin{array}{cc}
a^{\dagger}a+g(a^{\dagger}+a)  & \Delta \\
\Delta & a^{\dagger}a-g(a^{\dagger}+a) %
\end{array}%
\right).\label{Rabiha2}
\end{equation}

In recent years different methods have been proposed to construct analytic
solutions for the eigenspectrum of the quantum Rabi model.
In the Bargmann-Fock space of analytical functions \cite{Bargmann},
the quantum Rabi model can be mapped into two coupled
first-order ordinary differential equations for the wave function components $\psi_{1}$ and $\psi_{2}$
 \cite{Schw,Swain,ReikJ,Reika,Koc}.
Braak \cite{Braak, Braakb} used this approach to explicitly construct an analytic solution and
derived the conditions for determining the full energy spectrum
taking advantage of the $Z_2$ symmetry.
Braak's analytic solution was recovered in an alternative, more physical way using
a Bogoliubov transformation \cite{CWHLW2012}.
It was further found that the complete set of analytic solutions can be given in terms of a known special function --
the confluent Heun function~\cite{Ronveaux, Slavyanov}.
Different conditions for determining the full energy spectrum of the quantum Rabi model
were obtained in this way \cite{Zhong1,Maciejewskia}.
In the remainder of this section, we give an outline of the recent progress achieved using these methods.

\subsection{Solution via Bogoliubov transformation}

Soon after the solution obtained by Braak \cite{Braak} for the energy eigenspectrum,
the Bogoliubov transformation was applied~\cite{CWHLW2012} to give
the analytic solution of the quantum Rabi model.
The aim of the Bogoliubov transformation is to introduce a new bosonic operator
to remove the linear terms in the operators $a^{\dagger}$ and $a$ in hamiltonian (\ref{Rabiha2}).
Two different forms,
\begin{eqnarray}
A&=&a+g, \quad A^\dagger=a^\dagger+g,\\
B&=&a-g, \quad  B^\dagger=a^\dagger-g,
\end{eqnarray}
of such operators have been used.
In terms of the operators $A$ and $A^\dagger$, the hamiltonian $H_{\mathrm{R}}'$ becomes
\begin{equation}
H_{\mathrm{R}}'=\left(
\begin{array}{cc}
A^{\dagger}A-\alpha  & \Delta \\
\Delta & A^{\dagger}A-2g(A^{\dagger}+A)+\beta%
\end{array}%
\right),
\end{equation}
where $\alpha=g^2$ and $\beta=3g^2$.
The wave function components $\phi_1$ and $\phi_2$ are assumed to take the form
%
%
\begin{equation}
\left(
\begin{array}{c}
\phi_1 \\
\phi_2
\end{array}%
\right)=\left(
\begin{array}{c}
\sum_{n=0}^{\infty}\sqrt{n!}e_n|n_A\rangle \\
\sum_{n=0}^{\infty}\sqrt{n!}f_n|n_A\rangle
\end{array}%
\right),\label{Rabisola}
\end{equation}
where $e_n$ and $f_n$ are  the expansion coefficients and   $|n_A\rangle$ is
the extended coherent state defined by
\begin{equation}
|n_A\rangle=\frac{(A^{\dagger})^n}{\sqrt{n!}}|0_A\rangle,
\qquad |0_A\rangle=e^{-g(a^\dagger-a)}|0_a\rangle=e^{-g^2/2}|(-g)_a\rangle.
\end{equation}
Here $|0_a\rangle$ is the vacuum state and thus $|0_A\rangle$ is just the coherent state $|(-g)_a\rangle$.

Substitution into equation~(\ref{eigenvalue}) gives
\begin{eqnarray}
\sum_{n=0}^{\infty}(n-\alpha-E)\sqrt{n!}e_n|n_A\rangle+\Delta \sum_{n=0}^{\infty}\sqrt{n!}f_n|n_A\rangle=0,\\
\sum_{n=0}^{\infty}(n+\beta-E)\sqrt{n!}f_n|n_A\rangle+\Delta \sum_{n=0}^{\infty}\sqrt{n!}e_n|n_A\rangle \nonumber\\
-2g\left(\sum_{n=0}^{\infty}\sqrt{n}\sqrt{n!}f_n|(n-1)_A\rangle+\sum_{n=0}^{\infty}\sqrt{n+1}\sqrt{n!}f_n|(n+1)_A\rangle\right)=0.
\end{eqnarray}
Multiplying $\langle m_A|$ on both sides of the above equations then gives the relations
\begin{eqnarray}
\phantom{m} e_m&=&-\frac{\Delta}{n-\alpha-E}f_m,\\
m f_m&=&\Omega(m-1)f_{m-1}-f_{m-2}, \label{frec}
\end{eqnarray}
between the expansion coefficients $e_n$ and $f_n$, with
\begin{equation}
\Omega(m)=\frac{1}{2g}\left(m+\beta-E-\frac{\Delta^2}{m-\alpha-E}\right).
\end{equation}
Note that the coefficients $f_m$ in equation (\ref{frec}) obey a three-term recurrence relation.
The initial conditions may be chosen up to an overall normalisation as $f_0=1$ and $f_1=\Omega(0)$.

On the other hand, in terms of the operators $B$ and $B^\dagger$, the hamiltonian $H_{\mathrm{R}}'$ becomes
\begin{equation}
H_{\mathrm{R}}'=\left(
\begin{array}{cc}
B^{\dagger}B+2g(B^{\dagger}+B)+\beta'  & \Delta \\
\Delta & B^{\dagger}B-\alpha'%
\end{array}%
\right),
\end{equation}
where $\alpha'=\alpha=g^2$ and $\beta'=\beta=3g^2$.
In this case,  $\phi_1$ and $\phi_2$ can be written in the different form
\begin{equation}
\left(
\begin{array}{c}
\phi_1' \\
\phi_2'
\end{array}%
\right)=\left(
\begin{array}{c}
\sum_{n=0}^{\infty}(-1)^n\sqrt{n!}f_n'|n_B\rangle \\
\sum_{n=0}^{\infty}(-1)^n\sqrt{n!}e_n'|n_B\rangle
\end{array}%
\right).\label{Rabisolb}
\end{equation}
Following the above procedure gives
\begin{eqnarray}
\phantom{m} e_m'&=&-\frac{\Delta}{n-\alpha'-E}f_m',\\
m f_m'&=&\Omega'(m-1)f_{m-1}'-f_{m-2}',
\end{eqnarray}
with
\begin{equation}
\Omega'(m)=\frac{1}{2g} \left(m+\beta'-E-\frac{\Delta^2}{m-\alpha'-E} \right).
\end{equation}
Here the initial conditions may be chosen up to an overall normalisation as $f_0'=1$ and $f_1'=\Omega'(0)$.

Now an important issue is the relation between the two different forms of solutions (\ref{Rabisola}) and (\ref{Rabisolb}).
Physically, if this eigenvalue is not degenerate, they should represent the same state.
So they should only differ by a constant $r$, i.e.,
\begin{equation}
\left(
\begin{array}{c}
\phi_1 \\
\phi_2
\end{array}%
\right)=r\left(
\begin{array}{c}
\phi_1' \\
\phi_2'
\end{array}%
\right).
\end{equation}
Multiplying the vacuum state $\langle 0_a|$ on both sides of the above equations gives
\begin{eqnarray}
\sum_{n=0}^{\infty}e_ng^n=r\sum_{n=0}^{\infty}f_n'g^n,\\
\sum_{n=0}^{\infty}f_ng^n=r\sum_{n=0}^{\infty}e_n'g^n.
\end{eqnarray}
Eliminating the constant $r$ then yields the condition $G(E)=0$, where
\begin{equation}
G(E)= \sum_{n=0}^{\infty}\frac{\Delta \, f_ng^n}{n-\alpha-E}  \times \sum_{n=0}^{\infty}\frac{\Delta \, f_n'g^n}{n-\alpha'-E}
-\sum_{n=0}^{\infty}f_ng^n \times \sum_{n=0}^{\infty}f_n'g^n.
\end{equation}
Because $f_n$ and $f_n'$ satisfy the same recurrence relation under the conditions $\alpha=\alpha'=g^2$
and $f_0=f'_0=1$, it follows that $f_n=f_n'$.
The above condition simplifies to $G_{\pm}(x)=0$, with
\begin{equation}
G_{\pm}(x)=\sum_{n=0}^{\infty}f_n \left(1 \mp \frac{\Delta}{x-n} \right)g^n, \label{chenGfun}
\end{equation}
where we have set $E=x-g^2$.
It has been shown that the zeros of the $G$-functions give the energy spectrum \cite{Braak},
as we discuss further below.

\subsection{Braak's solution in Bargmann-Fock space}

Here one considers the Bargmann-Fock space $\mathcal{B}$ of analytical functions in a complex variable $z$.
In the Bargmann-Fock space,
the bosonic creation and annihilation operators have the form \cite{Bargmann,Schw,Swain,ReikJ,Reika,Koc}
\begin{equation}
a\rightarrow \frac{d}{d z}, \quad a^\dagger\rightarrow z.
\end{equation}
The hamiltonian $H_{\mathrm{R}}'$ thus becomes
\begin{equation}
H_{\mathrm{R}}'=\left(
\begin{array}{cc}
z\frac{d}{d z} +g(z+\frac{d}{d z}) & \Delta \\
\Delta  &  z\frac{d}{d z}-g(z+\frac{d}{d z}) %
\end{array}%
\right).\label{Bha1}
\end{equation}
In the Bargmann-Fock space, the eigenstate $\left|\phi\right\rangle$ of hamiltonian $H_{\mathrm{R}}'$ can be expressed as
\begin{equation}
\left|\phi\right\rangle=\left(
\begin{array}{c}
\phi_1(z) \\
\phi_2(z)
\end{array}%
\right).\label{Bsola}
\end{equation}

Using a Fulton-Gouterman transformation $\left|\varphi\right\rangle=U\left|\phi\right\rangle$, with
\begin{equation}
U=\frac{1}{\sqrt{2}}\left(
\begin{array}{cc}
1 & 1\\
T &  -T %
\end{array}%
\right),
\end{equation}
where $T$ is the reflection operator acting on elements $f(z)$ of $\mathcal{B}$: $Tf(z)=f(-z)$, gives
\begin{equation}
U^\dagger H_{\mathrm{R}}' U=\left(
\begin{array}{cc}
H_+ & 0\\
0 &  H_- %
\end{array}%
\right),
\end{equation}
with
\begin{equation}
H_{\pm}=z\frac{d}{d z} +g \left(z+\frac{d}{d z} \right) \pm \Delta T.
\end{equation}
This result implies that $H_+$($H_-$) acts in the subspace of positive (negative) parity $\mathcal{H_+}$ ($\mathcal{H_-}$).
In the subspace $\mathcal{H_+}$ with positive parity the Schr\"{o}dinger equation reads
\begin{equation}
z\frac{d}{d z}\varphi_1(z) +g \left(z+\frac{d}{d z} \right)\varphi_1(z)+\Delta  \varphi_1(-z)=E\varphi_1(z).
\end{equation}
This a functional differential equation in $z$.
Setting the notation $f_1(z)=\varphi_1(z)$ and $f_2(z)=\varphi_1(-z)$
one obtains a coupled system of first-order equations,
\begin{eqnarray}
(z+g)\frac{{df_1}}{{dz}} +(g z-E)f_1+ \Delta f_2=0,\\
(z-g)\frac{{df_2}}{{dz}} -(g z+E)f_2+ \Delta f_1=0.
\end{eqnarray}
%

To solve these equations we set $y=z+g$, $x=E+g^2$, $f_{1,2}(z)=e^{-gy+g^2}\chi_{1,2}(y)$ and obtain
\begin{eqnarray}
\phantom{(g-2)} y\frac{{d\chi_1}}{{d z}}&=&x\chi_1-\Delta \chi_2,\label{Bspacea}\\
(y-2g)\frac{{d\chi_2}}{{d z}}&=&(x-4g^2+2gy)\chi_2-\Delta \chi_1.\label{Bspaceb}
\end{eqnarray}
These coupled  equations are solved  by expanding $\chi_2(y)$ in a power series in $y$,
\begin{equation}
\chi_2(y)=\sum_{n=0}^{\infty}K_n(x)y^n.
\end{equation}
Then from equation (\ref{Bspacea}) we obtain
\begin{equation}
\chi_1(y)=\sum_{n=0}^{\infty}K_n(x)\frac{\Delta}{x-n}y^n.
\end{equation}
Equation (\ref{Bspaceb}) gives the three-term recurrence relation
\begin{equation}
nK_n=\Omega(n-1)K_{n-1}-K_{n-2}.
\label{Krec}
\end{equation}
Here the initial coefficients $K_0$ and $K_1$ are chosen to be  $K_0=1$ and $K_1=\Omega(0)$.
We note that  the coefficients $K_n$ have the same form as those of $f_n$ in the Bogoliubov transformation approach.

It therefore follows that there are two representations for $\varphi_1(z)$ in $\mathcal{H_+}$, namely
\begin{eqnarray}
\varphi_1^1(z)=f_1(z)=e^{-gz}\sum_{n=0}^{\infty}K_n(x)\frac{\Delta}{x-n}(z+g)^n, \\
\varphi_1^2(z)=f_2(-z)=e^{gz}\sum_{n=0}^{\infty}K_n(x)(g-z)^n.
\end{eqnarray}
Similarly there are two representations for $\varphi_2(z)$ in $\mathcal{H_-}$,
\begin{eqnarray}
\varphi_2^1(z)=e^{-gz}\sum_{n=0}^{\infty}K_n(x)\frac{-\Delta}{x-n}(z+g)^n, \\
\varphi_2^2(z)=e^{gz}\sum_{n=0}^{\infty}K_n(x)(g-z)^n.
\end{eqnarray}
From the two representations for $\varphi_{1}$ and $\varphi_{2}$ we have
\begin{eqnarray}
G_+(x,z)&=&\varphi_1^2(z)-\varphi_1^1(z)=0, \\ G_-(x,z)&=&\varphi_2^2(z)-\varphi_2^1(z)=0.
\end{eqnarray}

\begin{figure}[t]
\begin{center}
\includegraphics[width=0.7\columnwidth]{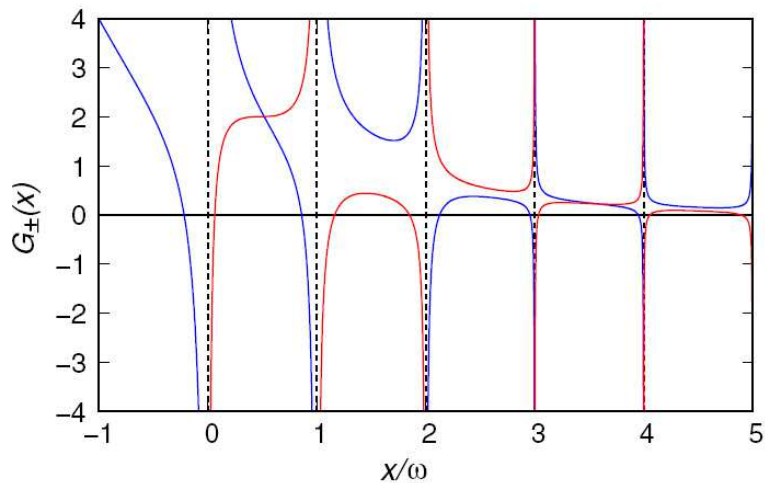}
\caption{The quantum Rabi model functions $G_+(x)$ (red lines) and $G_-(x)$ (blue lines)
as functions of the variable $x/\omega$ in the interval $[-1,5]$ for parameter values
$g/\omega=0.7$ and $\Delta/\omega=0.4$. The regular parts of the energy eigenspectrum follow from the
conditions $G_\pm(x)=0$. Poles at the values $x=n\omega$ correspond to the exceptional two-fold degenerate Judd points.
Reproduced with permission from~\cite{Braak}.}\label{gfun3}
\end{center}
\end{figure}

The conditions $G_\pm(x,z)=0$ hold in the whole complex plane if and only if $x=E+g^2$
corresponds to a point in the energy spectrum of the quantum Rabi model.
From a mathematical point of view, iff $x=E+g^2$ is a point in the energy spectrum of the quantum Rabi model,
both $\varphi_1^{1,2}(z)$ and $\varphi_2^{1,2}(z)$ are analytic in the whole complex plane.
For $z=0$ we have the simplified form
\begin{equation}
G_{\pm}(x)=\sum_{n=0}^{\infty}K_n(x)\left(1\mp \frac{\Delta}{x-n}\right)g^n. \label{BraakGfun}
\end{equation}
This function is plotted in figure \ref{gfun3} and discussed further in Section 3.
It can be written directly in terms of confluent Heun functions~\cite{Braak_review}.

\subsection{Solution in terms of confluent Heun functions}

The confluent Heun function~\cite{Ronveaux, Slavyanov} appears naturally
in the analytic solution for the wave function components in the Bargmann-Fock space.
It follows from equations (\ref{Bha1}) and (\ref{Bsola}) that\footnote{The solution outlined here
in terms of confluent Heun functions follows that given by the present authors~\cite{Zhong1}.
The reader is also referred to the compact treatment given in~\cite{Maciejewskia}, in which
the confluent Heun functions appear naturally around the relevant singularities.
As far as we are aware, the first mention of confluent Heun functions in the context of the quantum Rabi model was
by D. Braak (see \cite{CWHLW2012}).}
\begin{eqnarray}
(z+g)\frac{{d\phi_1}}{{dz}} +(g z-E)\phi_1+ \Delta \phi_2=0,\label{Bceqa}\\
(z-g)\frac{{d\phi_2}}{{dz}} -(g z+E)\phi_2+ \Delta \phi_1=0.\label{Bceqb}
\end{eqnarray}
We note that the same equations can be obtained via a different method \cite{Wu1,Wu2}.
Eliminating $\phi_2$ and $\phi_1$ from these two equations gives
second-order differential equations for $\phi_1(z)$ and $\phi_2(z)$ of the form
\begin{eqnarray}
\frac{{d^2\phi_1}}{{dz^2}}+p_1(z)\frac{{d\phi_1}}{{dz}}+q_1(z)\phi_1=0,\label{Bseqa}\\
\frac{{d^2\phi_2}}{{dz^2}}+p_2(z)\frac{{d\phi_2}}{{dz}}+q_2(z)\phi_2=0,\label{Bseqb}
\end{eqnarray}
with
\begin{eqnarray}
p_{1,2}(z) &=& \frac{(1-2E-2g^2)z+\mp g}{z^2-g^2},\nonumber\\
q_{1,2}(z) &=&\frac{-g^2 z^2+\mp g
z+E^2-g^2-\Delta^2}{z^2-g^2}.\nonumber
\end{eqnarray}

Using an appropriate variable transformation, these equations for $\phi_{1}$ and $\phi_{2}$
can be transformed into the confluent Heun equation,
such that their solutions follow in terms of confluent Heun functions \cite{Zhong1}.
The two different types of solutions
\begin{eqnarray}
\phi_1^1(z)&=&e^{-g z}\textrm{HC}\left(\alpha_1,\beta_1,\gamma_1,\delta_1,\eta_1,\frac{g-z}{2g}\right),\label{af1a}\\
\phi_2^1(z)&=&\frac{\Delta }{E+g^2}e^{-g
z}\textrm{HC}\left(\alpha_2,\beta_2,\gamma_2,\delta_2,\eta_2,\frac{g-z}{2g}\right),\label{af2a}
\end{eqnarray}
and
\begin{eqnarray}
\phi_1^2(z)&=&\frac{\Delta }{E+g^2}e^{g
z}\textrm{HC}\left(\alpha_2,\beta_2,\gamma_2,\delta_2,\eta_2,\frac{g+z}{2g}\right),\label{af1b}\\
\phi_2^2(z)&=&e^{g
z}\textrm{HC}\left(\alpha_1,\beta_1,\gamma_1,\delta_1,\eta_1,\frac{g+z}{2g}\right).
\label{af2b}
\end{eqnarray}
follow from equations (\ref{Bseqa}) and (\ref{Bseqb}) (details are given in Appendix A).
Here
\begin{equation}
\textrm{HC}(\alpha,\beta,\gamma,\delta,\eta,x)=\sum_{n=0}^{\infty} h_nx^n
\label{HCdef}
\end{equation}
is the confluent Heun function.
The coefficients $h_n$ are defined from the three-term recurrence relation
\begin{equation}
A_n h_n=B_n h_{n-1}+C_n h_{n-2}, \qquad n\geq 1,
\label{recdef}
\end{equation}
with initial conditions $h_0=1$ and $h_{-1}=0$ and
\begin{eqnarray}
A_n&=&1+\beta/n,\\
B_n&=&1+(\beta+\gamma-\alpha-1)/n+[\eta-\beta/2+(\gamma-\alpha)(\beta-1)/2]/n^2,\\
C_n&=&[\delta+\alpha(\beta+\gamma)/2+\alpha(n-1)]/n^2.
\end{eqnarray}
The various parameters appearing in the confluent Heun functions are
$\alpha_2=\alpha_1=4g^2$, $\beta_2=\gamma_1=-(E+g^2)$,
$\gamma_2=\beta_1=-(E+g^2+1)$, $\delta_2=-\delta_1=2g^2$
and $\eta_2=\eta_1+\delta_1=-3g^4/2-(3+2E)g^2/2+(E^2+E-\Delta^2+1)/2$.

The two equations (\ref{Bceqa}) and (\ref{Bceqb}) have $Z_2$ symmetry.
This can be seen by replacing $z$ by $-z$ in the equations, giving
\begin{eqnarray}
(z-g)\frac{{d\phi_1(-z)}}{{d z}}-(gz+E)\phi_1(-z)+\Delta \phi_2(-z)=0,\\
(z+g)\frac{{d\phi_2(-z)}}{{d z}}+(gz-E)\phi_2(-z)+\Delta \phi_1(-z)=0.
\end{eqnarray}
It is clearly seen that if $(\phi_1(z),\phi_2(z))$ are solutions of equations (\ref{Bceqa}) and (\ref{Bceqb})
then $(\phi_2(-z),\phi_1(-z))$ are also solutions.
If the energies are not degenerate, then
\begin{equation}
\left(
\begin{array}{c}
\phi_1(z) \\
\phi_2(z)
\end{array}%
\right)=C\left(
\begin{array}{c}
\phi_2(-z) \\
\phi_1(-z)
\end{array}%
\right).
\end{equation}
This leads to the two possible values $C\pm 1$, with
$C=1$ and $C=-1$ corresponding to the symmetric and anti-symmetric solutions, respectively.
From the two sets of different solutions $\phi_1^{1,2}$ and $\phi_2^{1,2}$
the symmetric and anti-symmetric solutions,
\begin{eqnarray}
\phi_1^+(z)&=&\phi_1^1(z)+\phi_1^2(z),\\
\phi_2^+(z)&=&\phi_2^1(z)+\phi_2^2(z),
\end{eqnarray}
and
\begin{eqnarray}
\phi_1^-(z)&=&\phi_1^1(z)-\phi_1^2(z),\\
\phi_2^-(z)&=&\phi_2^1(z)-\phi_2^2(z),
\end{eqnarray}
are constructed.
They satisfy the relations $\phi_{1,2}^+(z)=\phi_{2,1}^+(-z)$ and $\phi_{1,2}^-(z)=-\phi_{2,1}^-(-z)$.

Naturally, $\phi_{1,2}^+$ and $\phi_{1,2}^{-}$ are required to satisfy the two coupled equations
(\ref{Bceqa}) and (\ref{Bceqb}).
This leads to two different relations
\begin{eqnarray}
K^{\pm}(E, z)&:=& \e^{-gz} \, G_1^{\pm}(E,z)\mp \Delta \, \e^{gz} \, G_2^{\pm}(E,z) \nonumber\\
&:=& \e^{-gz} \, G_3^{\pm}(E,z)\pm \frac{\Delta}{E+g^2} \, \e^{gz} \, G_4^{\pm}(E,z) \nonumber\\
&=&0,\label{kplus}
\end{eqnarray}
where
\begin{eqnarray}
G_1^{\pm}(E, z)&=&F_1(E,z)\pm \frac{\Delta}{E+g^2} \, \e^{2gz} \, F_4(E, z), \label{Gfun1}\\
G_2^{\pm}(E,z)&=&F_3(E, z)\pm \frac{\Delta}{E+g^2} \, \e^{-2gz} \, F_2(E,z),\\
G_3^{\pm}(E, z)&=&F_1(E,z)\mp \Delta \, \e^{2gz} \, F_3(E, z), \\
G_4^{\pm}(E,z)&=&F_4(E, z)\mp \Delta \, \e^{-2gz} \, F_2(E,z) \label{Gfun2},
\end{eqnarray}
with
\begin{eqnarray}
F_1(E,z)&=&(E+g^2)\, \textrm{HC}\left(\alpha_1,\beta_1,\gamma_1,\delta_1,\eta_1,\frac{g-z}{2g}\right)\nonumber\\
&&+\frac{g+z}{2g} \textrm{HC}' \left(\alpha_1,\beta_2,\gamma_1,\delta_1,\eta_1,\frac{g-z}{2g}\right),\nonumber\\
F_2(E,z) &=&\textrm{HC}\left(\alpha_2,\beta_2,\gamma_2,\delta_2,\eta_2,\frac{g-z}{2g}\right),\nonumber\\
F_3(E,z)&=&\textrm{HC}\left(\alpha_1,\beta_1,\gamma_1,\delta_1,\eta_1,\frac{g+z}{2g}\right),\nonumber\\
F_4(E,z)&=&(E-g^2-2g z)\textrm{HC}\left(\alpha_2,\beta_2,\gamma_2,\delta_2,\eta_2,\frac{g+z}{2g}\right)\nonumber\\
&&-\frac{g+z}{2g} \textrm{HC}'\left(\alpha_2,\beta_2,\gamma_2,\delta_2,\eta_2,\frac{g+z}{2g}\right).\nonumber
\end{eqnarray}
Here $\textrm{HC}'(\alpha,\beta,\gamma,\delta,\eta,x)$
denotes the derivative of the confluent Heun function (\ref{HCdef}) with respect to $x$.
From the relation $K^{\pm}(E,z)=0$ we can also obtain two sets of sufficient conditions,
namely the pair of weaker conditions
\begin{equation}
G_1^{\pm}(E,z)=G_2^\pm(E,z)=0 \quad \textrm{and} \quad G_3^\pm(E,z)=G_4^\pm(E,z)=0. \label{weaker}
\end{equation}
It is clear that if these conditions are satisfied, then $K^{\pm}(E,z)=0$.
It has been shown that $G_{1,2,3,4}^{\pm}=0$  and Braak's $G_\pm=0$
give the same result for the energy spectrum of the quantum Rabi model \cite{Zhong1},
as discussed further in Section 3.1.
Such conditions hold for $z$ in the range $|(g-z)/2g|<1$ and  $|(g+z)/2g|<1$, where the
confluent Heun functions are convergent.

It should be noted that the Braak $G$-function approach essentially involves gluing together two local series solutions
to obtain a global solution which is entire.
The motivation for this section has been to present this line of approach to the end in terms of confluent Heun functions.
In this sense the solution to the eigenspectrum problem of the quantum Rabi model is seen to involve known functions.

\subsection{Integrability}

There has been some recent discussion concerning the issue of quantum integrability
and the Rabi model~\cite{Braak,Moroz2013,BZ,Moroz2016}.
One underlying factor is that, unlike the single clearcut definition of classical integrability,
there are various definitions of quantum integrability~\cite{CM,Larson2013}.
Arguably for one-dimensional quantum systems, the most appropriate definition is Yang-Baxter integrability.
However, the concept of Yang-Baxter integrability applies to many-body systems.
One could take the strict view that the quantum Rabi model, comprised of one
qubit interacting with a single mode of a quantised light field, does not
constitute a many-body system, at least in the Yang-Baxter sense.
Yet from another perspective, exactly solved models are known to go hand-in-hand with integrable systems.
Given that analytic solutions have been obtained for the quantum Rabi model of the type reviewed in this article,
it is natural to expect some kind of underlying integrability.
To this end, Braak proposed a phenomenological level-labelling criterion of quantum integrability \cite{Braak}, which is satisfied
by the quantum Rabi model.
Despite previous attempts, the quantum Rabi model does not appear to be Yang-Baxter integrable,
at least in the usual form and in the full parameter space~\cite{BZ}.
However, an inkling that this might be possible has been provided by the connection between the
confluent Heun equation and Painlev\'e V~\cite{Painleve}.
In this way the energy spectrum of the quantum Rabi model can be obtained in terms of Painlev\'e V transcendents~\cite{Painleve},
however the ramifications of this result are still to be fully understood.
Also on the question of integrability, it was pointed out recently~\cite{Moroz2016} that
a numerical study of two-dimensional patterns of quantum invariants in the
anisotropic Rabi model~\cite{QI1,QI2} implies that the quantum Rabi model is not integrable.

The quantum Rabi model has been claimed to be a quasi-exactly solved model \cite{Moroz2013,ZhangY}.
The concept of quasi-exact solvability applies to a quantum system for which only part of the eigenspectrum
can be derived algebraically~\cite{Turb,BD,quasi}.
For the quantum Rabi model this is the exceptional part of the eigenspectrum made up of the isolated exact Judd points,
which can be derived algebraically, among a number of different approaches, as discussed further below.
It seems however, that for all intents and purposes the quantum Rabi model
can be regarded as an exactly solved model,
since an analytic solution has been obtained and applies to all parts of the eigenspectrum.

\section{Energy spectrum and dynamics}

Over the past decades various methods have been used to compute the energy spectrum of the quantum Rabi model,
or equivalently, the energy spectrum of the JC model beyond the RWA.
The problem of finding the energy eigenvalues is reduced to the diagonalisation of an infinite tridiagonal matrix~\cite{Graham},
which can be done numerically by truncation of the matrix to finite order.
Some other approaches are series expansions~\cite{Feng1999} and continued fractions~\cite{Feng2001,Ziegler,Braakcf,Moroz2012,Moroz2014a,Moroz2014b}.
Various other computational schemes have been discussed (see, e.g., \cite{Irish2014} and references therein)
with the necessary aim to be effective in the ultrastrong and deep strong coupling regimes.
A method known as the generalised RWA has been used~\cite{FKU,I2007,Albert2011,ZhangCZ2013,YZLCJ, ZhangChen2015,ZCHC} to calculate physical quantities of interest.
Another approach uses an analytic approximation based on an unitary transformation~\cite{Zheng2012,Zheng2015}.
Variational approaches have also been applied~\cite{var1,var2},
most recently in a polaron-antipolaron context~\cite{var3,var4}.
We mention also a continued fraction and three-term recurrence relation approach which has been developed to calculate the energy spectrum
using an $F$-function~\cite{Moroz2013,Moroz2012,Moroz2014a,Moroz2014b}.
This is different to Braak's $G$-function, but works in a similar fashion.

Our emphasis here is on the energy spectrum obtained via the analytic approach through $G$-functions.

\subsection{Energy spectrum}

As stated already, the energy spectrum of the quantum Rabi model includes both regular and exceptional parts.
These parts of the energy spectrum can be clearly defined in terms of the variable
$x=E/\omega + g^2/\omega^2$~\cite{Braak,Braak_review}.
Specifically, the regular spectrum consists of the values $E_n$ for which $x_n$ is not a non-negative integer.
The exceptional spectrum consists of the values $E_n$ for which $x_n$ is a non-negative integer.
The regular parts of the energy spectrum correspond to the non-degenerate parts determined by the zeros of the $G$-functions,
\begin{equation}
G_{\pm}(E,z)=0, \qquad G^{\pm}_{1,2,3,4}(E,z)=0,
\end{equation}
given in equations (\ref{BraakGfun}) and (\ref{Gfun1})--(\ref{Gfun2}).

\begin{figure}[tb]
\begin{center}
 \includegraphics[width=0.9\columnwidth]{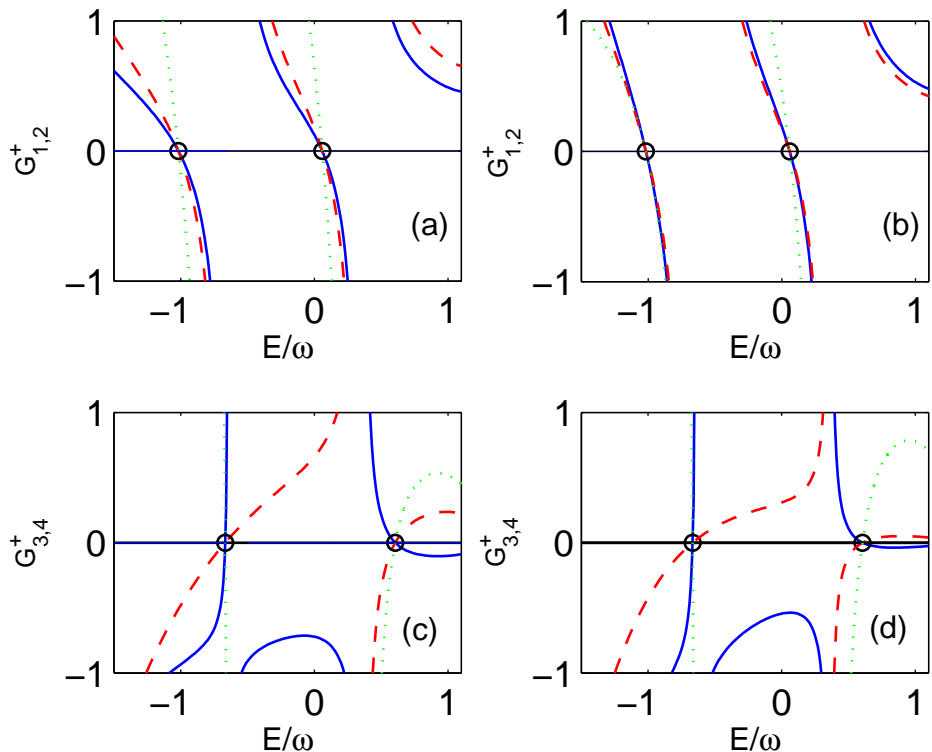}
\caption{$G_{1,2,3,4}^{+}(E, z)$ as functions of the energy $E/\omega$ for $z=0$ ((a) and (c)) and $z=0.3$ ((b) and (d)).
The parameter values are $\omega=1$, $\Delta/\omega=0.7$ and $g/\omega=0.8$.
The solid lines are for $G_{1,3}^+(E,z)$ and the dashed lines for $G_{2,4}^+(E,z)$.
Here the zero axis defines the conditions $G_{1,2,3,4}^{+}(E, z)=0$.
The circles thus correspond to the allowed energies. For comparison, also shown are Braak's  \cite{Braak} functions
$G_-(E,z)$ (dotted lines in (a) and (b)) and $G_+(E,z)$ (dotted lines in (c) and (d)).
It is seen that $G_{1,2}^{+}(E, z)$ and $G_-(E,z)$ have the same roots.
Similarly for $G_{3,4}^{+}(E, z)$ and $G_+(E,z)$.
Reproduced with permission from~\cite{Zhong1}.}\label{gfunfig1}
\end{center}
\end{figure}

\begin{figure}[tb]
\begin{center}
\includegraphics[width=0.9\columnwidth]{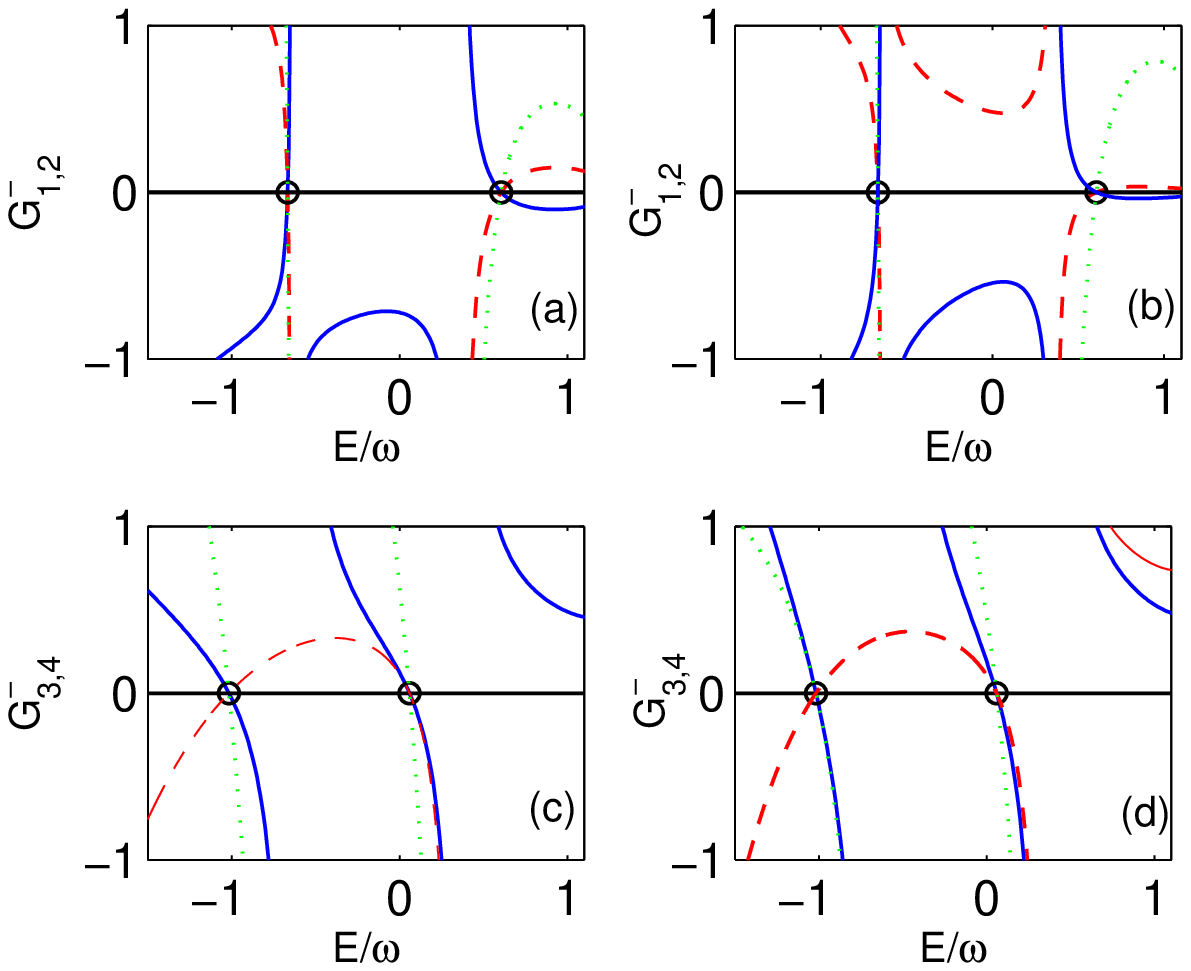}
\caption{$G_{1,2,3,4}^{-}(E, z)$ as functions of the energy $E/\omega$ with for $z=0$ ((a) and (c)) and $z=0.3$ ((b) and (d)).
The parameter values are $\omega=1$, $\Delta/\omega=0.7$ and $g/\omega=0.8$.
The solid lines are for  $G_{1,3}^-(E,z)$ and the dashed lines for  $G_{2,4}^-(E,z)$.
Here the zero axis defines the conditions $G_{1,2,3,4}^{-}(E, z)=0$.
The circles thus correspond to the allowed energies. For comparison, also shown are Braak's  \cite{Braak} functions
$G_+(E,z)$ (dotted lines in (a) and (b)) and $G_-(E,z)$ (dotted lines in (c) and (d)).
It is seen  that $G_{3,4}^-(E, z)$ and $G_-(E,z)$ have the same roots.
Similarly for $G_{1,2}^{-}(E, z)$ and $G_+(E,z)$.
Reproduced with permission from~\cite{Zhong1}.}\label{gfunfig2}
\end{center}
\end{figure}

\begin{figure}[tb]
\begin{center}
\includegraphics[width=0.8\columnwidth]{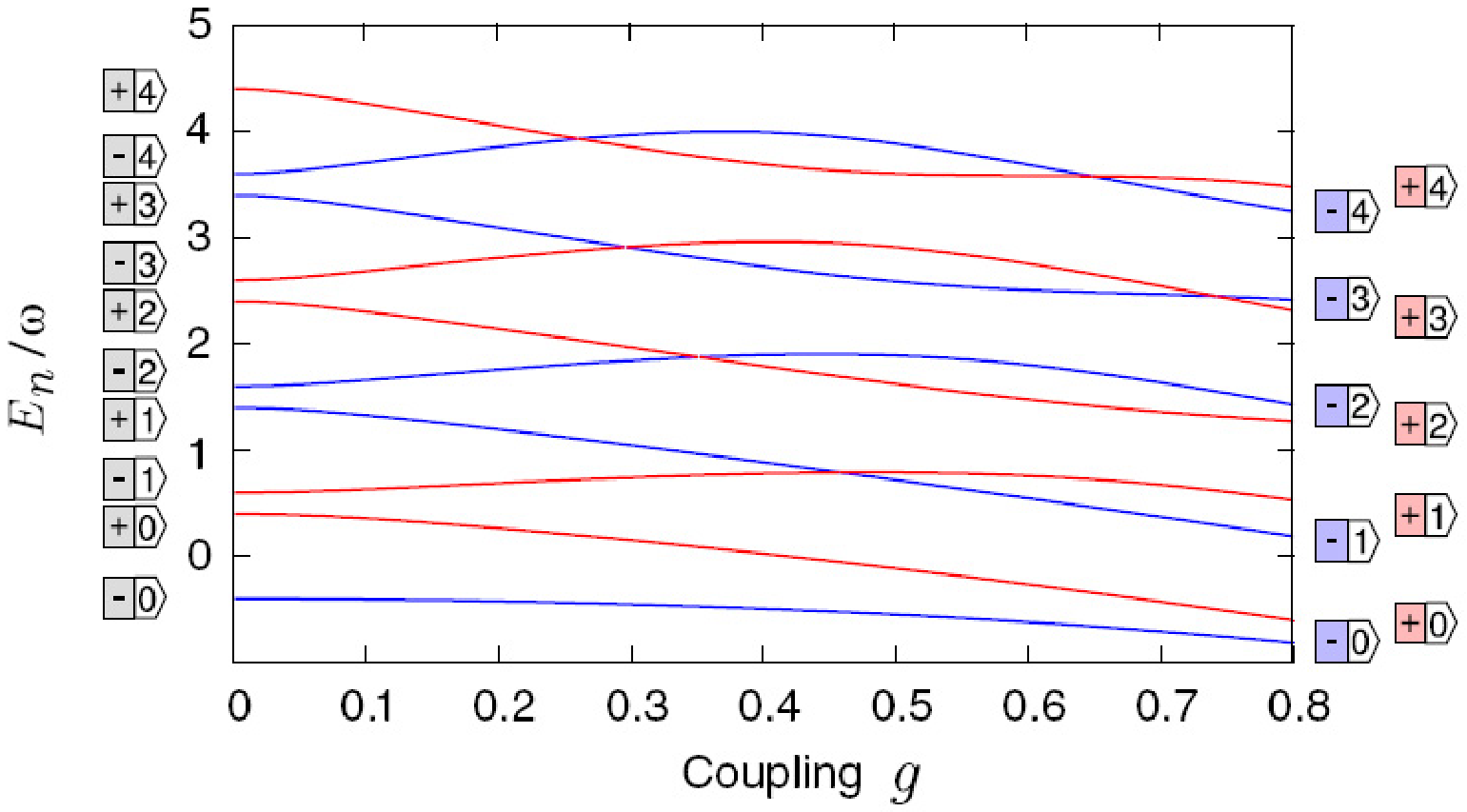}
\caption{Energy  spectrum of the quantum Rabi model as a function of the coupling $g$
for parameter values $\Delta=0.4$ and $\omega=1$. The red and blue lines correspond to positive
and negative  parities, respectively.  On the left side the states with $g=0$ are labeled by
$|\pm,n\rangle$, with $+$ or $-$ corresponding to the two-level system and $n=0,1,2,...$
to the eigenstates of the bosonic mode. Reproduced with permission from~\cite{Braak}}\label{Rabiena}
\end{center}
\end{figure}

It has been shown that $G_+$, $G^{+}_{3,4}$, and $G^{-}_{1,2}$ have the same zeros,
as can be seen in figure \ref{gfunfig1}.
In addition,  $G_-$, $G^{+}_{1,2}$, and $G^{-}_{3,4}$ have the same zeros,
as shown in figure \ref{gfunfig2}.\footnote{Note there is a typo in the figure 2 caption
in Ref.~\cite{Zhong1}: $G_-(E,z)$ and $G_+(E,z)$ should be interchanged.}
There are no level crossings within each parity subspace.
This allows a unique labelling of each state by a pair of quantum numbers: the parity $p=\pm 1$
and $n=0,1,2, \ldots$ which are the $n$th zeros of the $G$-functions
corresponding to the eigenstates of the bosonic mode.
This labelling is used in figure \ref{Rabiena}, which shows the low-lying  energy spectrum as a function of the coupling $g$.

Although the analytic solutions $\phi_1^1(z)$ ($\phi_2^1$(z)) and $\phi_1^2(z)$ ($\phi_2^2(z)$) given above
seem to be of different forms, they are actually two linearly dependent solutions
when $\beta_{1}$ ($\beta_2$) and $\gamma_1$ ($\gamma_{2}$) are not integers.
For two linearly dependent solutions their Wronskian must be zero, i.e.,
\begin{equation}
W_1(E,z) := \phi_{1}^{2}\frac{d\phi_{1}^{1}}{dz}-\phi_{1}^{1}\frac{d\phi_{1}^{2}}{dz}=0,
\end{equation}
and
\begin{equation}
W_2(E,z) := \phi_{2}^{2}\frac{d\phi_{2}^{1}}{dz}-\phi_{2}^{1}\frac{d\phi_{2}^{2}}{dz}=0.
\end{equation}
Since $\phi_2^1(-z)=\phi_1^2(z)$ and $\phi_1^1(-z)=\phi_2^2(z)$, it follows that $W_1(E,-z)=W_2(E,z)$.
The conditions $W_{1,2}(E,z)=0$ hold for arbitrary values of $z$ if and
only if $E$ corresponds to an eigenvalue in the energy spectrum of
the generalised quantum Rabi model~\cite{Braak}.
Therefore the energy spectrum of the quantum Rabi model can also be obtained
with the Wronskian method~\cite{Zhong1,Maciejewskia}.
Indeed, the above Wronskians can be written in terms of the confluent Heun functions.

\subsection{Exceptional points}

The exceptional part of the energy spectrum corresponds to poles of the $G$-functions, with energy
\begin{equation}
{E_n}/{\omega}=n- {g^2}/{\omega^2}.
\end{equation}
At these exceptional points the parameters $g/\omega$ and $\Delta/\omega$ satisfy relations among themselves given by the
condition
\begin{equation}
K_n(n\omega)=0,
\label{Kcon}
\end{equation}
where $K_n$ satisfies the three-term recurrence relation (\ref{Krec}).
This condition (\ref{Kcon}) defines a series of constraint polynomials with increasing $n$.
The exceptional points correspond to the level crossings between the two parity subspaces,
some of which can be seen in figure~\ref{Rabiena}.
These points were first noticed by Judd~\cite{Judd} and are known as Judd or Juddian solutions, sometimes
also called isolated exact solutions.
They have been discussed by a number of authors and can be obtained via different
approaches (see, e.g., \cite{Judd,ReikJ,Koc,kus2,Reikb,Emaryb,ReikE,kus}.
For example, more recently it was shown that the exceptional points and related constraint polynomials
can be obtained from a system of coupled Bethe Ansatz type equations~\cite{ZhangY}.
From an analysis of the constraint polynomials, Ku\'s \cite{kus} was able to prove that there are
$n-k$ crossings points in the energy spectrum for each value of $n$ for $\Delta/\omega$ in the range
$k < \Delta/\omega < k+1$.

The exceptional solutions can be obtained in a systematic way from the analytic solution obtained
in terms of confluent Heun functions.
This is because, under certain parameter conditions,
the confluent Heun functions can be terminated as a polynomial~\cite{Ronveaux,Slavyanov,Fiz2010}.
For the confluent Heun function (\ref{HCdef}) the set of conditions is
\begin{eqnarray}
h_{N+1}&=&0, \label{conditionb}\\
\quad \quad \delta&=&-(N+(\gamma+\beta+2)/2)\alpha,\label{conditiona}
\end{eqnarray}
for integer $N\geq 0$.
Under these conditions, the coefficients $h_n$ vanish for $n>N$.
The Heun function thus truncates to $N$ terms.
Applying the conditions (\ref{conditionb}) and (\ref{conditiona}) in this way leads to Judd's isolated exact solutions,
with a finite set of recurrence relations following from (\ref{recdef}).
The coefficients are given by
\begin{eqnarray}
A_n&=&n(n-1-N),\\
B_n&=&(1-n+N)^2 - 4(n-1)g^2/\omega^2 - \Delta^2/\omega^2,\\
C_n&=&4(n-2-N)g^2/\omega^2.
\end{eqnarray}

For the confluent Heun functions appearing in $\phi_{1,2}^1(z)$ to be truncated as a finite series, we require that
$E/\omega=N_1+1-g^2/\omega^2$ with the coefficient $h_{N_1+1}=0$ in $\phi_1^1(z)$ and
$E/\omega=N_2-g^2/\omega^2$ with the coefficient $h_{N_2+1}=0$ in $\phi_2^1(z)$.
By way of illustration, consider the case $N_2=0$ and $N_1=1$, for which the energy eigenvalue
\begin{equation}
{E}/{\omega}=1-{g^2}/{\omega^2},\label{exam1}
\end{equation}
subject to the parameter relation
\begin{equation}
{\Delta^2}/{\omega^2}+{4g^2}/{\omega^2}
-1=0. \label{exam1con}
\end{equation}
The analytic solutions truncate to
\begin{eqnarray}
\phi_1^{1}(z)&=&\left(1-2g^2 +2gz \right) e^{-g z},\\
\phi_2^{1}(z)&=&\Delta e^{-g z}.
\end{eqnarray}

In the second set of analytical solutions, $\phi_{1}^{2}(z)$ and $\phi_{2}^{2}(z)$,
we have $E/\omega=N_1+1-g^2/\omega^2$ with the coefficient $h_{N_1+1}=0$ in $\phi_1^2(z)$ and
$E/\omega=N_2-g^2/\omega^2$ with the coefficient $h_{N_2+1}=0$ in $\phi_2^2(z)$.
In the case of $N_1=0$ and $N_2=0$, we have again (\ref{exam1}) and (\ref{exam1con}), with now the solutions
\begin{eqnarray}
\phi_1^{2}(z)&=& \Delta e^{g z} ,\\
\phi_2^{2}(z)&=&\left(1-2g^2 -2gz \right) e^{g z}.
\end{eqnarray}
The two sets of solutions are clearly degenerate.
From these, the symmetric and anti-symmetric solutions,
\begin{eqnarray}
\phi_1^{\pm}(z)&=& \e^{-gz} \, (1-2g^2+2g z)\pm\Delta \, \e^{gz},\\
\phi_2^{\pm}(z)&=& \Delta \, \e^{-gz} \pm \e^{gz} \, (1-2g^2-2g z),
\end{eqnarray}
can readily be constructed.
They satisfy $\phi_{1,2}^+(z)=\phi_{2,1}^+(-z)$ and $\phi_{1,2}^-(z)=-\phi_{2,1}^-(-z)$.
Importantly, from these exact symmetric and anti-symmetric solutions, it follows that
\begin{equation}
K^\pm(E,z)=0, \quad G_{1,2,3,4}^\pm(E,z)\neq 0.
\end{equation}
This example thus gives an explicit demonstration that the weaker conditions $G_{1,2,3,4}^\pm(E,z)=0$
are only applicable for the regular part of the energy spectrum of the quantum Rabi model.

The two-fold degenerate Judd points discussed so far are not the only exceptional points in the energy spectrum.
The other, non-degenerate, set of exceptional points have also been discussed \cite{Braak_review,ReikJ,ReikE,Maciejewskib,Li2}.
The system parameters do not satisfy a constraint polynomial at these points,
but other conditions can be derived.

\subsection{Level spacing statistics}

\begin{figure}[t]
\begin{center}
\includegraphics[width=0.8\columnwidth]{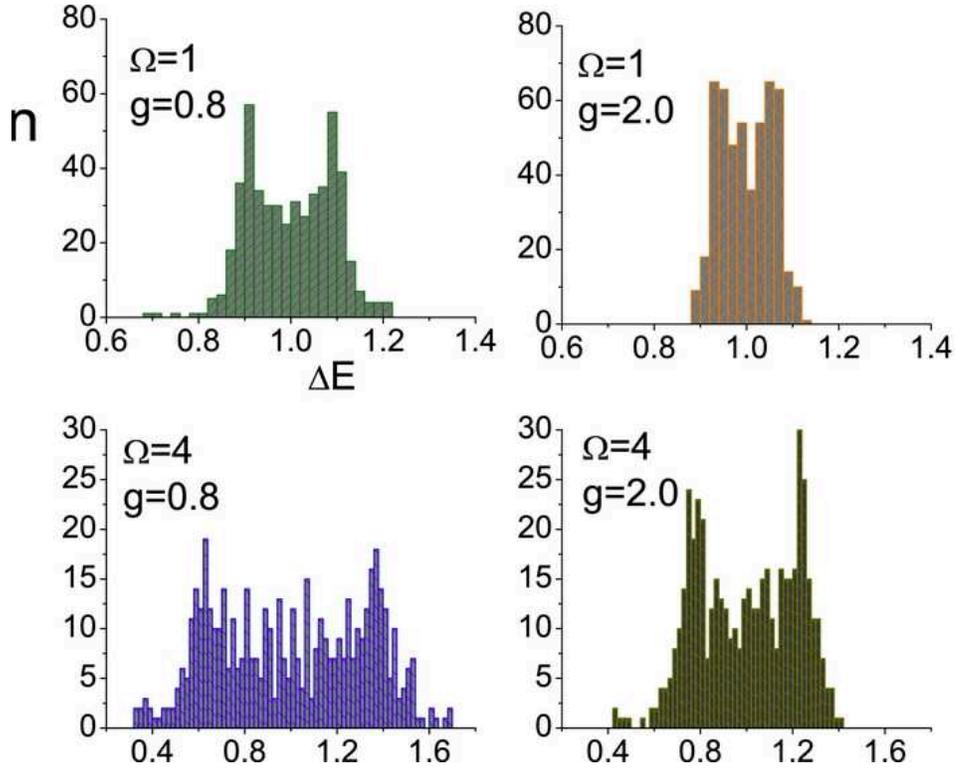}
\end{center}
\caption{Distribution of spacings between neighbouring energy levels for the quantum Rabi model in the
+ parity subspace. The parameter values $\Delta=\Omega$ and $g$ are as indicated.
The lowest 501 levels are considered.
Reproduced with permission from \cite{WL2013}.}\label{enstab}
\end{figure}

An important physical quantity to investigate the statistical properties of the energy spectrum
is the distribution of nearest-neighbour spacings defined by $\Delta E_n=E_{n+1}-E_n$.
Early computations of the nearest-neighbour spacing distribution \cite{Graham,Kus1985}
indicated that neither Poissonian nor Wigner-Dyson distributions provide a good description for the results.
Recall that Poissonian distributions are characteristic of integrable systems, whereas
Wigner-Dyson distributions are characteristic of chaotic systems.
More recent and extensive results \cite{WL2013} are shown in figure \ref{enstab}.
The conclusion is that the distribution has a two-peak structure, with peaks located at $\Delta E< 1$ and $\Delta E > 1$.
The peaks become narrower and move closer together as $g$ increases.
The peaks become wider and separate from each other as $\Delta$ increases.
Ku\'s~\cite{Kus1985} also identified two factors which ``confine" the overall shape of the distribution
for higher energy levels, particularly if one plots $\Delta E_n$ vs $n$.
These are the Judd points and the $g \to \infty$ (integrable) limit, where the eigenvalues are also
two-fold degenerate and given by $E_n/\omega = n - g^2/\omega^2$, $n = 0, 1, 2, \ldots$.

Some insights on the spectrum statistics can also be gained from the perspective
of the $G$-functions \cite{Braak}.
We know that the zeros of $G_\pm$ correspond to the regular parts of the energy spectrum.
If the parameters deviate from the conditions for the exceptional solutions to hold,
the distribution of the zeros of the $G$-functions is regular.
Specifically, the number of eigenvalues in each interval $[n\omega,(n+1)\omega)$ is restricted to be $0,1$, or $2$
for a given parity, as can be seen in figure \ref{gfun3}.
Moreover, it is believed that an interval $[n\omega,(n+1)\omega)$ with 2 roots of $G_\pm=0$
can only be adjacent to an interval with 1 or 0 roots.
Similarly an empty interval can never be adjacent to another empty interval.
There are no level crossings within a parity subspace.

\subsection{Dynamics}

\begin{figure}[t]
\begin{center}
\includegraphics[width=0.9\columnwidth]{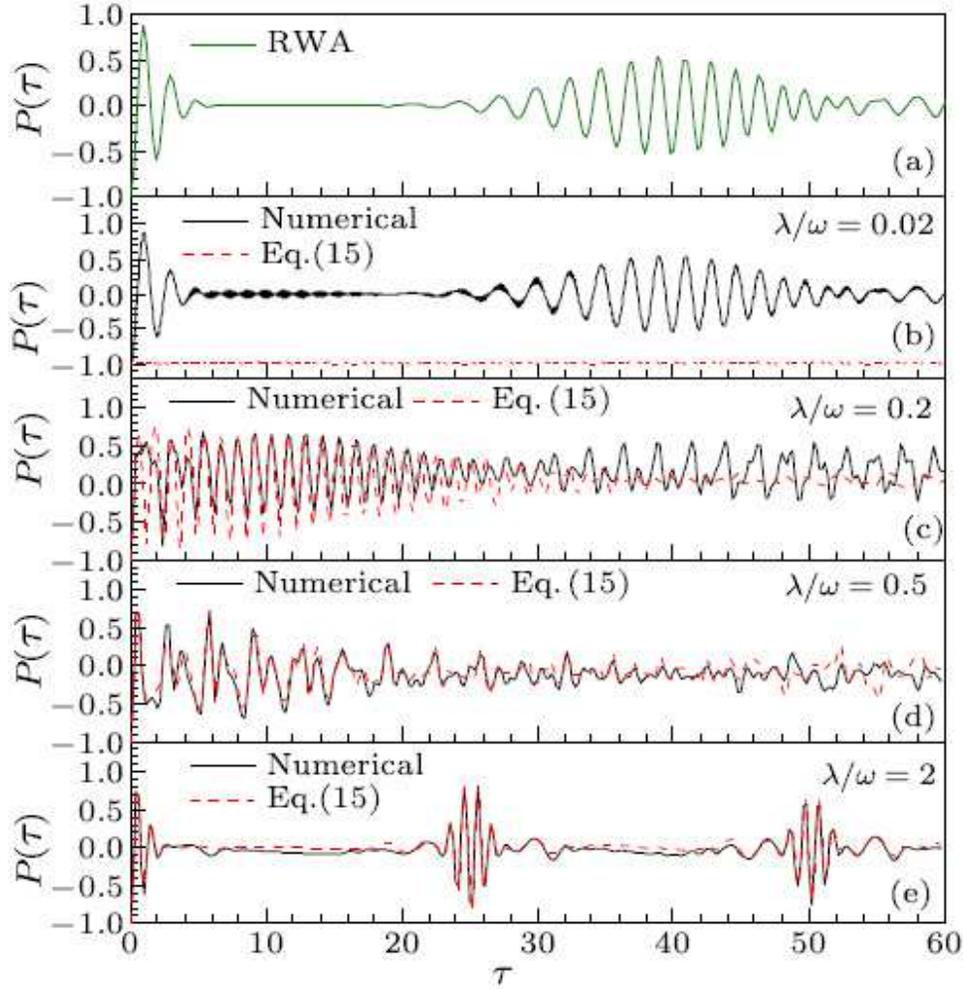}
\caption{Time evolution of the population inversion $P(t)$ (solid curves) in the dimensionless time scale $\tau=2gt$
 (a) with  RWA, and for different
coupling strength $g/\omega=\lambda/\omega=0.02$ (b), $0.2$ (c), $0.5$ (d) and $2.0$ (e)
obtained by numerical diagonalization (black solid line) and the analytical
solution in equation (\ref{scre}) (dashed red line).
Here the initial mean photons $\alpha= \sqrt{10}$ and $2\Delta=\omega$.
Reproduced with permission from~\cite{ZCZ2013}.}\label{Rabidyna}
\end{center}
\end{figure}

We now turn to the dynamical behaviour of the quantum Rabi model.
The state vector  $|\psi(t)\rangle$ at any time $t$ is a linear combination
\begin{equation}
|\psi(t)\rangle=\sum_{n=0}^{\infty}c^{\uparrow}_n(t)|\!\uparrow,n
\rangle+\sum_{n=0}^{\infty}c^{\downarrow}_n(t)|\!\downarrow,n \rangle
\end{equation}
of the basis states
$| \! \uparrow,n \rangle$ and $| \! \downarrow,n \rangle$ satisfying
\begin{equation}
\sigma_z|\! \uparrow,n\rangle=|\!\uparrow,n\rangle, \quad
\sigma_z|\!\downarrow,n\rangle=-|\!\uparrow,n\rangle,
\end{equation}
and
\begin{equation}
a^{\dagger}a|\!\uparrow,n\rangle=n|\!\uparrow,n\rangle, \quad
a^{\dagger}a|\! \downarrow,n\rangle=n|\!\downarrow,n\rangle.
\end{equation}
Substituting this expansion into the time-dependent Schr\"{o}dinger equation gives
\begin{eqnarray}
\mathrm{i} \frac{{d}}{{dt}} c^{\uparrow}_n(t) &=&
(n\omega+\Delta) c^{\uparrow}_n(t)+
g(\sqrt{n}c^{\downarrow}_{n-1}(t)+\sqrt{n+1}c^{\downarrow}_{n+1}(t)),\label{ceqa}\\
\mathrm{i} \frac{{d}}{{dt}} c^{\downarrow}_n(t)
&=&(n\omega-\Delta)
c^{\downarrow}_n(t)+
g(\sqrt{n}c^{\uparrow}_{n-1}(t)+\sqrt{n+1}c^{\uparrow}_{n+1}(t)).\label{ceqb}
\end{eqnarray}
It is observed that $|\! \uparrow, n  \rangle$ is coupled to the states
$|\! \downarrow, n+1  \rangle$ and $|\! \downarrow, n-1  \rangle$.
Similarly $|\! \downarrow, n  \rangle$ is coupled to
$|\! \uparrow, n+1  \rangle$ and $|\! \uparrow, n-1  \rangle$.
This coupling is related to the parity operator (\ref{parity}),
which is a conserved quantity, with eigenvalues $p=\pm 1$.
It follows that the dynamics moves inside the Hilbert space consisting of two unconnected subspaces
labelled by their parity,
\begin{eqnarray}
| \! \downarrow, 0  \rangle \leftrightarrow | \! \uparrow, 1  \rangle \leftrightarrow | \! \downarrow,
2  \rangle \leftrightarrow | \! \uparrow, 3  \rangle \leftrightarrow \ldots \quad p=+1,\nonumber\\
| \! \uparrow, 0  \rangle \leftrightarrow | \! \downarrow, 1  \rangle \leftrightarrow | \! \uparrow,
2  \rangle \leftrightarrow | \! \downarrow, 3  \rangle \leftrightarrow \ldots \quad p=-1.
\end{eqnarray}

The initial state is assumed to be  a coherent state in the upper level $| \! \uparrow \rangle$,
\begin{equation}
|\psi(0)\rangle=| \! \uparrow,\alpha \rangle=\sum_{n}\frac{e^{-|\alpha|^2/2}\alpha^n}{\sqrt{n!}}| \! \uparrow,n \rangle.
\end{equation}
A key quantity is the time evolution of the population inversion $P(t)$ defined by
\begin{equation}
P(t)=\langle \sigma_z\rangle=\langle\psi(t)|\sigma_z|\psi(t)\rangle=\sum_{n}(|c^{\uparrow}_{n}|^2-|c^{\downarrow}_{n}|^2).
\end{equation}
In the weak coupling regime,  $P(t)$ can be obtained exactly under the RWA.
For the particular case $2\Delta=\omega$,
\begin{equation}
P(t)=\sum_{n}|c^{\uparrow}_{n}(0)|^2\cos(2 \sqrt{n+1} g t).
\end{equation}
In this weak coupling regime, the inversion $P(t)$ shows the phenomena of collapse and revival (CR).
The CR of the population inversion was first predicted by Eberly {\em et al.} \cite{Eberly}
and later observed experimentally for a single Rydberg atom in a cavity \cite{Rempe}.
The effect of the CRT on the dynamics of the inversion has been studied numerically in a wide
range of $g/\omega$~\cite{ZCZ2013,Nagel}.
As $g/\omega$ is increased the CR phenomena disappears but reemerges in the deep strong coupling regime
(see figure \ref{Rabidyna}).
In the large  coupling regime, there is also an analytical result \cite{ZCZ2013},
\begin{equation}
P(t)=-\sum_{nm}\frac{a_n[b_{n,m}\cos(\Delta E^+_{n,m})-\mu_{n,m}\cos(\Delta E^-_{n,m})]}{4\sqrt{a^2_n+b^2_{n,m}+\mu^2_{n,m}}},\label{scre}
\end{equation}
where  $a_n=f_n^{+2}+f_n^{-2}$,
$b_{n,m}=D_{nm}(2g/\omega)f_n^+f_m^+$, $\mu_{n,m}=D_{nm}(2g/\omega)f_n^-f_m^-$, $f_n^{\pm}=e^{-(g/\omega-\alpha)^2}[1\pm(-1)^n]$,$D_{nm}(x)=e^{-x^2/2}\sum_{i=0}^{\min [n,m]}(-1)^i\sqrt{n!m!}x^{n+m-2i}/i!(m-i)!(n-i)!$, $\Delta E^{\pm}_{n,m}=E_n^{\pm}-E_m^{\pm}$, and $E_m^{\pm}=\omega(m-g^2/\omega^2)\mp(-\Delta)D_{mm}(2g/\omega)$.  In the strong coupling regime, the terms $\Delta E^{\pm}_{n,m}$
can be further approximated to be $\omega(n-m)$.
Like the above result obtained for the RWA,
this result involves the summation of $\cos$ terms, thereby leading to the appearance of CR.

\begin{figure}[tb]
\begin{center}
\includegraphics[width=0.8\columnwidth]{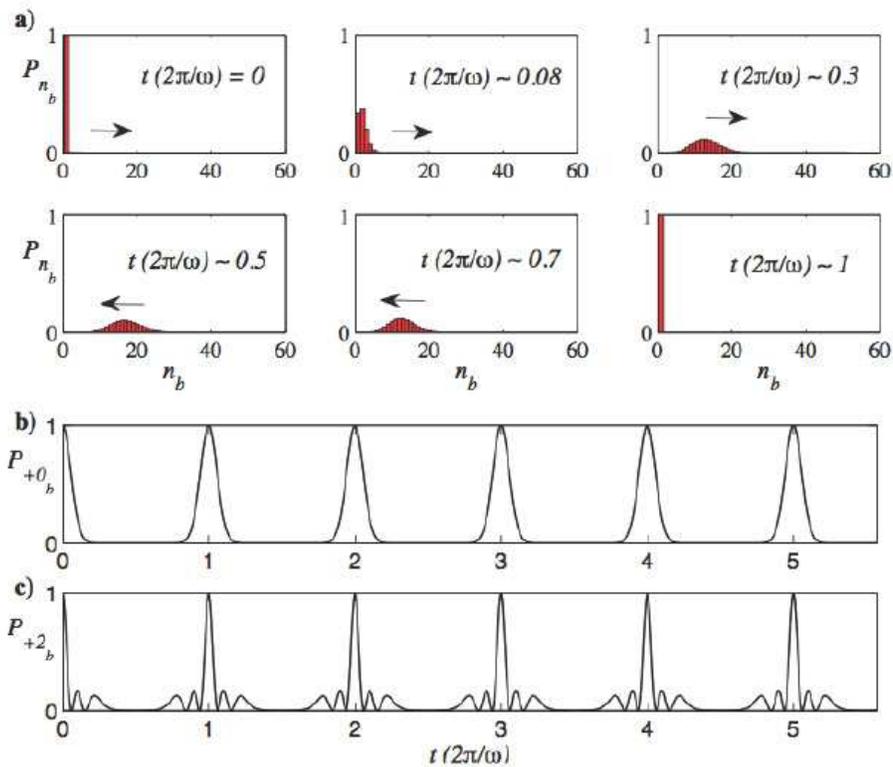}
\caption{(a)-(b) Round trip of a photon number wavepacket and CRs in the deep strong coupling regime of the quantum Rabi model with initial
state $|+,0_b\rangle=| \!\downarrow,0_a\rangle$. c) CRs with secondary peaks
due to counter propagating photon number wavepackets with the
initial state $|+,2_b\rangle=| \!\downarrow,2_a\rangle$.
For all cases $\Delta=0$ and $g/\omega=2$. Reproduced with permission from~\cite{Casanova}.}\label{Rabidynb}
\end{center}
\end{figure}

An intuitive physical framework has been provided for understanding the reappearance of the  CR ~\cite{Casanova}.
In terms of the parity operator $\Pi$ and $b=\sigma_x b$, the Rabi hamiltonian $H_R$ can be written as
\begin{equation}
H_R=b^\dagger b+g(b^\dagger+b)-\Delta(-1)^{b^\dagger b}\Pi.
\end{equation}
Here one can introduce  the parity basis $|p,n_b\rangle$ with $b^\dagger b |p,n_b\rangle=n_b|p,n_b\rangle$ and $\Pi |p,n_b\rangle=p|p,n_b\rangle$.
The resulting hamiltonian describes a perturbed harmonic oscillator where the last term behaves as an energy shift.
If  the initial state is $|\psi(0)\rangle=|+,0_b\rangle=|\downarrow,0_a\rangle$,
the particular case $\Delta=0$ is exactly solvable.
The wave function at any time $t$ is given as
\begin{equation}
|\psi(t)\rangle=e^{ig^2t}e^{-ig^2\sin(t)}|+,\alpha(t)\rangle,
\end{equation}
where $\alpha(t)=g(e^{-it}-1)$ is the amplitude of the coherent state.
The revival probability of the initial state is
\begin{equation}
P_{+0_b}=|\langle \psi(0)|\psi(t)\rangle|^2=e^{-|\alpha(t)|^2},
\end{equation}
exhibiting periodic CR, as shown in figure \ref{Rabidynb}.

Comparison of CR phenomena between the JC and quantum Rabi models, among other quantities,
has also been made~\cite{Larson2007,Naderi}.
We mention here also recent simulations of oscillator tunneling dynamics~\cite{Irish2014},
universal dynamics under slow quenches~\cite{plenio},
and the dynamics under drive and dissipation~\cite{LeHur} in the quantum Rabi model.
For related and other dynamical aspects of the JC and quantum Rabi models,
see also~\cite{PC2008,FL2009a,FL2009b,FL2011,SMF2016,FLS2016}.

\subsection{Berry phase}

\begin{figure}[tb]
\begin{center}
\includegraphics[width=1.0\columnwidth]{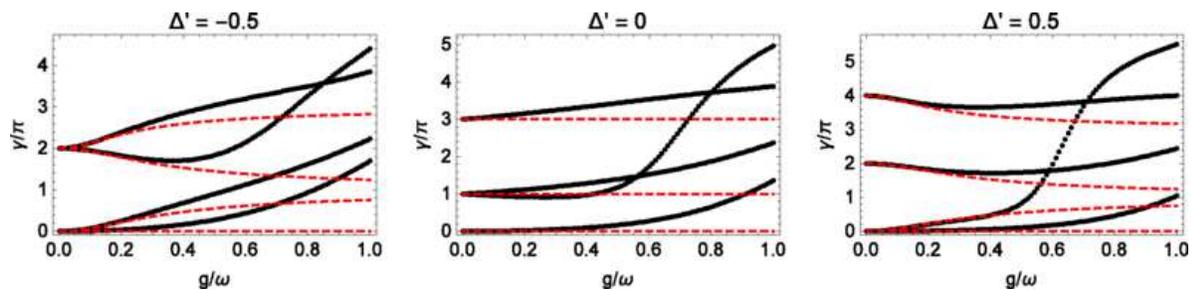}
\caption{Geometric phase $\gamma$ (in units of $\pi$) associated to eigenvectors of the quantum Rabi hamiltonian
as a function of $g/\omega$. The parameter $\Delta' = (2 \Delta - \omega)/\omega$ with $\Delta'=0$ the resonant case.
Dashed lines correspond to phases associated to eigenvectors of the JC hamiltonian.
Reproduced with permission from~\cite{CZ2016}.}\label{phases}
\end{center}
\end{figure}

Geometric phases have been predicted to appear when a two-level system interacts with a quantised field,
including the vacuum state.
Results have been reported regarding the existence of a Berry phase in the quantum
Rabi model induced by the unitary transformation
$U(\varphi) = \exp(-\mathrm{i} \varphi \, a^\dagger a)$~\cite{LFW2011,DL2013, WWL2015, CZ2016}.
This was in contrast with a previous result where it was argued
that the appearance of such Berry phases is an artifact of the RWA \cite{Larson}.
This issue appears to have been settled, with a
nonvanishing Berry phase appearing in both the JC and Rabi models.
This Berry phase is given by
\begin{equation}
\gamma_n = 2\pi \langle \psi_n | a^\dagger a | \psi_n \rangle,
\end{equation}
with $| \psi_n \rangle$ the $n$th eigenstate of the hamiltonian considered.
The phases $\gamma_n$ have been calculated numerically as a function of $g/\omega$ (see figure~\ref{phases}).

Other aspects of the geometric curvature and phase in the quantum Rabi model have also
been discussed recently~\cite{MHGZ2015}.

\section{Generalised quantum Rabi models}

After Braak's analytic solution of the quantum Rabi model,
analytic solutions for the full eigenspectrum of
various known generalisations were found.
These more general models are the
\begin{itemize}
\item asymmetric quantum Rabi model \cite{Braak, CWHLW2012,Maciejewskib,Zhong2},
\item anisotropic quantum Rabi model \cite{X2014,Tomka,Shen,Zhang},
\item two-photon quantum Rabi model \cite{CWHLW2012,T2012,Maciejewski,Zhang2015,Cui,Peng2013,T2015,DXBC},
\item two-mode quantum Rabi model \cite{Cui,DuanEPL15,Chilingaryan2015},
\item two-qubit quantum Rabi model \cite{ Peng2013,Peng2015,Peng2014,Duan, Wang,Chilingaryan2013},
\item Dicke model \cite{Braak2013,He2015},
\item $N$-state quantum Rabi model \cite{ZhangY2014}.
\end{itemize}
In addition the quantum Rabi model has been solved in similar fashion
with an additional  nonlinear coupling term $\sigma_z \, a^\dagger a$ between the atom and the cavity~\cite{Maciejewskib}.

In this section we give an outline of the solutions obtained for the first three models on this list,
namely the asymmetric quantum Rabi model, the
anisotropic quantum Rabi model and the two-photon quantum Rabi model.
The structure of the solution for the two-mode quantum Rabi model shares a common structure with the
one- and two-photon quantum Rabi models~\cite{DuanEPL15}.
Two-mode squeezed states are important because several devices produce correlated light at two frequencies.
Some isolated exact solutions have been found for the
two-mode quantum Rabi model~\cite{ZhangY,Zhang2015,Chilingaryan2015} and
the two-qubit quantum Rabi model~\cite{Peng2015,Peng2014,Chilingaryan2013,MHZ2015}.
The problem of two-qubits has potential applications in quantum information technology because various quantum information resources, such as quantum entanglement and quantum discord,
can be easily stored in two qubits in a common cavity~\cite{Romero}.
Likewise the Dicke model \cite{Dicke}, which is the extension of the quantum Rabi model to $N$ qubits, with each qubit
interacting with the same single mode of the cavity, is of immense practical importance.
The analytic solutions obtained so far for the Dicke model are for $N=3$~\cite{Braak2013}
and also for general $N$~\cite{He2015}, although much work remains to be done in the general case.

\subsection{Asymmetric quantum Rabi model}

A generalised Rabi model is the  asymmetric quantum Rabi model
\begin{equation}
H_{\mathrm{R}}^\epsilon=\omega a^{\dagger}a+g\sigma_x(a^{\dagger}+a)+\Delta
\sigma_z+\epsilon\sigma_x.
\end{equation}
This model is also known as the biased quantum Rabi model.
It has been studied numerically in the context of thermalisation~\cite{Larson2013}.
Due to the presence of the last term in the hamiltonian, the $Z_2$ symmetry is broken.
However, $H_{\mathrm{R}}^\epsilon$ can be embedded into a larger system possessing a $Z_2$ symmetry.
In the Bargmann-Fock space, the model can be solved in the extended Hilbert space  \cite{Braak}.
This model can also be solved via the Bogoliubov transformation \cite{CWHLW2012}.
The solution can be obtained by replacing $\alpha=g^2$ and $\beta=3g^2$
by $\alpha=g^2-\epsilon$ and $\beta=3g^2-\epsilon$ in equation (\ref{Rabisola}),
$\alpha'=g^2$ and $\beta'=3g^2$ by $\alpha'=g^2+\epsilon$ and $\beta'=3g^2+\epsilon$ in equation (\ref{Rabisolb}).
It has been shown that like the quantum Rabi model,  the $n$th eigenvalue $E_n$ of the regular parts of the energy spectrum
is determined by the $n$th zero of the $G$-function
\begin{equation}
G_\epsilon=\Delta^2\bar{R}^+(x)\bar{R}^-(x)-R^+(x)R^-(x),
\end{equation}
where
\begin{equation}
R^{\pm}=\sum_{n=0}^\infty K_n^{\pm}g^n, \bar{R}^{\pm}=\sum_{n=0}^\infty \frac{K_n^{\pm}}{x-n\pm \epsilon}g^n.
\end{equation}
Here we have set $f_n=K_n^-$,  $f_n'=K_n^+$ and $E=x-g^2$.
In addition, in the Bargmann-Fock space, the solutions can also be expressed in terms of confluent Heun functions \cite{Maciejewskib,Zhong2}.
The two sets of solutions are now
\begin{eqnarray}
\phi_1^1(z)&=&e^{-g z}\textrm{HC} \left(\alpha_1,\beta_1,\gamma_1,\delta_1,\eta_1,\frac{g-z}{2g}\right),\\
\phi_2^1(z)&=&\frac{\Delta e^{-g
z}}{E+g^2+\epsilon}\textrm{HC} \left(\alpha_2,\beta_2,\gamma_2,\delta_2,\eta_2,\frac{g-z}{2g}\right),
\end{eqnarray}
and
\begin{eqnarray}
\phi_1^2(z)&=&\frac{\Delta e^{g
z}}{E+g^2-\epsilon}\textrm{HC} \left(\alpha_1,\gamma_1,\beta_1,-\delta_1,\eta_1+\delta_1,\frac{g+z}{2g}\right),\\
\phi_2^2(z)&=&e^{g
z}\textrm{HC} \left(\alpha_2,\gamma_2,\beta_2,-\delta_2,\eta_2+\delta_2,\frac{g+z}{2g}\right).
\end{eqnarray}
In $\phi_1^{1,2}(z)$,  the parameter values are
$\alpha_1=4g^2$, $\beta_1=-(E+\epsilon+g^2+1)$, $\gamma_1=-(E-\epsilon+g^2)$,
$\delta_1=-2(1-2\epsilon)g^2$
and $\eta_1=-3g^4/2+(1-2E-4\epsilon)g^2/2+(E^2+E-\epsilon^2+\epsilon-2\Delta^2+1)/2$.
In $\phi_2^{1,2}(z)$,  $\alpha_2=4g^2$, $\beta_2=-(E+\epsilon+g^2)$, $\gamma_2=-(E-\epsilon+g^2+1)$,
$\delta_2=2(1+2\epsilon)g^2$
and $\eta_2=-3g^4/2-(3+2E+4\epsilon)g^2/2+(E^2+E-\epsilon^2-\epsilon-2\Delta^2+1)/2$.

It has been shown that although the analytic solutions
$\phi_1^1(z)$ ($\phi_2^1(z)$) and $\phi_1^2(z)$ ($\phi_2^2(z)$) appear to  have
different forms, they are actually two linearly dependent solutions
when $\beta_{1}$ ($\beta_2$) and $\gamma_1$ ($\gamma_{2}$) are non-integer.
Since they are linearly dependent, their Wronskian may be used to construct the conditions
\begin{eqnarray}
W_1^\epsilon(E,z) &:=&
\phi_{1}^{2}\frac{d\phi_{1}^{1}}{dz}-\phi_{1}^{1}\frac{d\phi_{1}^{2}}{dz}=0,\\
W_2^\epsilon(E,z) &:=&
\phi_{2}^{2}\frac{d\phi_{2}^{1}}{dz}-\phi_{2}^{1}\frac{d\phi_{2}^{2}}{dz}=0,
\end{eqnarray}
for the energy spectrum, from which the regular parts of the spectrum can be obtained.

The exceptional parts of the energy spectrum can be given from the conditions for
truncation of the confluent Heun functions, and take the form~\cite{Zhong2}
\begin{equation}
{E_n^{\pm}}/{\omega}=n-{g^2}/{\omega^2}\pm {\epsilon}/{\omega},
\end{equation}
where for the purposes of this discussion we have restored $\omega$.
The energy separation between exceptional points of the same integer $n$ is determined by the parity-breaking term
\begin{equation}
E^{+}_{n} - E^{-}_{n} = 2 \epsilon.
\end{equation}
This means that two exceptional points with the same integer $n$ gradually separate when $\epsilon$ increases.
For some specific values of $\epsilon$, exceptional points with different $n$ may form degenerate points of energy.
For example, if the condition
\begin{equation}
{\epsilon}/{\omega}=\frac{1}{2}(n_2-n_1)
\end{equation}
is satisfied, we may have $E_{n_1}^+=E_{n_2}^-$.
This is to say that the two exceptional points form a two-fold degenerate energy point when
$\epsilon$ is an integer multiple of $\frac{1}{2}\omega$ \cite{Braak,Zhong2}.
Constraint polynomials for exceptional points have also been discussed for this model~\cite{Zhong2,Li}.
The results obtained by Ku\'s \cite{kus} for the number of level crossings for the $\epsilon=0$ case can be generalised~\cite{Li}.
For a given value of $N$ there are $N$ level crossings for $0 < \Delta/\omega < \sqrt{1+2\epsilon/\omega}$,
reducing to $N-k$ crossing points for $\Delta$ in the range
\begin{equation}
\sqrt{k^2 + 2 k \epsilon/ \omega} < \Delta/\omega < \sqrt{(k+1)^2 + 2(k+1) \epsilon/ \omega} \,.
\label{cross}
\end{equation}
The class of exceptional points not satisfying a constraint polynomial have also been discussed for this model~\cite{Li2}.

\begin{figure}[tb]
\begin{center}
\includegraphics[width=0.6\columnwidth]{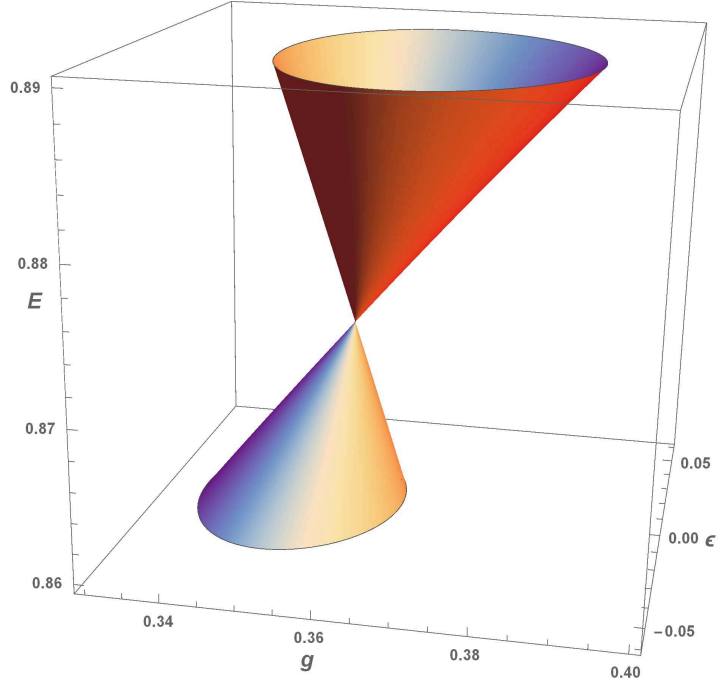}
\caption{Elementary cone centred at $\epsilon=0$ in the energy spectrum
of the asymmetric quantum Rabi model in the $g$-$\epsilon$ plane
for parameter values $\Delta=0.7$ and $\omega=1$.
Reproduced with permission from~\cite{BLZ}.}\label{cone}
\end{center}
\end{figure}

It has been shown that the effect of the bias parameter $\epsilon$ is to induce a conical intersection point
in the energy spectrum at each of the
two-fold degenerate Judd points located at $\epsilon =0$~\cite{BLZ}.
Conical intersection points also occur for $\epsilon$ an integer multiple of $\frac{1}{2}\omega$.
Their precise location, and the energy landscape in general, have been explored in the $g$-$\epsilon$ plane.
A typical conical intersection is shown in figure~\ref{cone}.
It remains to investigate the influence of this landscape on the physical properties of the model.
For example, geometric phases associated to trajectories encircling such conical intersection points
are expected to be nonvanishing.

\subsection{Anisotropic quantum Rabi model}

The anisotropic quantum Rabi model is defined by the hamiltonian
\begin{equation}
H =\omega a^{\dagger}a+\Delta
\sigma_z+g(a^{\dagger}\sigma^{-}+a\sigma^{+})+\lambda g
(a^{\dagger}\sigma^{+}+a\sigma^{-}).
\end{equation}
Here the two rotating terms have different couplings, with
$\lambda=0$ reducing to the JC model and $\lambda=1$ the quantum Rabi model.
Different couplings of this kind appear to have first been introduced in the context of the more general Dicke model~\cite{Hioe}.
The anisotropic quantum Rabi model can be realised in different physical systems \cite{X2014}.
In particular, it has been shown that the model can be used to fit experimental data for
superconducting circuits in the strongly coupled regime \cite{Mooij,X2014}.
The anisotropic quantum Rabi model has recently been investigated in the context of
quantum state engineering in hybrid open quantum systems~\cite{JLS2016}.

The eigenspectrum of the anisotropic quantum Rabi model has been obtained via the $G$-function approach by
various authors \cite{X2014,Shen,Zhang}.
Isolated exact solutions have also been discussed \cite{Tomka}, along with a connection of the general
hamiltonian to differential operators of Dunkl type~\cite{Moroz2016}.
In the Bargmann-Fock space,  the eigenstate $\left|\psi\right\rangle$ can be expressed as
\begin{equation}
\left|\psi\right\rangle=\left(
\begin{array}{c}
\psi_1 \\
\psi_2
\end{array}%
\right),
\end{equation}
and thus from $H|\psi\rangle=E|\psi\rangle$ follow the coupled equations
\begin{eqnarray}
z\frac{{d\psi_1}}{{d z}}+g\frac{{d\psi_2}}{{d z}}+\lambda g z\psi_2+\Delta \psi_1 =E\psi_1, \label{diffa} \\
\lambda g \frac{{d\psi_1}}{{d z}}+z\frac{{d\psi_2}}{{d z}}+g
z\psi_1-\Delta \psi_2 =E\psi_2\label{diffb}.
\end{eqnarray}
The linear combinations $\phi_1=\psi_1+\psi_2$ and $\phi_2=\psi_1-\psi_2$ then lead to the
equations for $\phi_1(z)$ and $\phi_2(z)$
\begin{eqnarray}
&&\left(z+\frac{1+\lambda}{2}g\right)\frac{{d\phi_1}}{{d z}}+\frac{1+\lambda}{2}gz\phi_1-\frac{1-\lambda}{2}g\frac{{d\phi_2}}{{d z}}
\nonumber\\
&& \qquad \qquad \qquad \qquad \qquad \qquad  +\frac{1-\lambda}{2}gz\phi_2+\Delta \phi_2 =E\phi_1,\\
&&\left(z-\frac{1+\lambda}{2}g\right)\frac{{d\phi_2}}{{dz}}-\frac{1+\lambda}{2}gz\phi_2+\frac{1-\lambda}{2}g\frac{{d\phi_1}}{{d
z}} \nonumber\\
&& \qquad \qquad \qquad \qquad \qquad \qquad  -\frac{1-\lambda}{2}gz\phi_1+\Delta \phi_1 =E\phi_2.
\end{eqnarray}
It is important to note that this model has a $Z_2$ symmetry of the form $\phi_1(-z)=\pm \phi_2(z)$, which
plays a crucial role in solving the model.

The above equations are solved by introducing the transformation
\begin{eqnarray}
\phi_1=\frac{(1+\sqrt{\lambda})\varphi_1-(1-\sqrt{\lambda})\varphi_2}{1+\lambda},\\
\phi_2=\frac{(1-\sqrt{\lambda})\varphi_1+(1+\sqrt{\lambda})\varphi_2}{1+\lambda},
\end{eqnarray}
which leads to the different set of coupled equations
\begin{eqnarray}
&&(z+g\sqrt{\lambda})\frac{{d\varphi_1}}{{d z}}+g\sqrt{\lambda}z\varphi_1+\frac{1-\lambda}{1+\lambda}\Delta \varphi_1 \nonumber \\
&& \qquad \qquad \qquad -g(1-\lambda)\frac{{d\varphi_2}}{{d z}}+\frac{2\sqrt{\lambda}}{1+\lambda}\Delta \varphi_2 =E\varphi_1, \\
&&(z-g\sqrt{\lambda})\frac{{d\varphi_2}}{{dz}}-g\sqrt{\lambda}z\varphi_2-\frac{1-\lambda}{1+\lambda}\Delta\varphi_2 \nonumber\\
&& \qquad \qquad \qquad -(1-\lambda)gz\varphi_1+\frac{2\sqrt{\lambda}}{1+\lambda}\Delta
\varphi_1 =E\varphi_2.
\end{eqnarray}
At this stage, the method suggested by Braak \cite{Braak} is followed by setting
$y=z+g\sqrt{\lambda}$,
$x=E+g^2\lambda-\frac{1-\lambda}{1+\lambda}\Delta$,
$\alpha=-(1-\lambda)g^2\sqrt{\lambda}-\frac{2\sqrt{\lambda}}{1+\lambda}\Delta$ and
$\varphi_{1,2}=f_{1,2}\exp[-g\sqrt{\lambda}z]$.
In such a way follow the coupled equations
\begin{eqnarray}
y\frac{{df_1}}{{d y}}-xf_1=\alpha f_2+(1-\lambda)g\frac{{df_2}}{{d y}},\label{aRabia}\\
(y-2g\sqrt{\lambda})\frac{{df_2}}{{dz}}+\left(-2g\sqrt{\lambda}y+4g^2\lambda-2\frac{1-\lambda}{1+\lambda}\Delta-x\right)f_2 \nonumber \\
\qquad \qquad =\alpha f_1+(1-\lambda)gyf_1.
\label{aRabib}
\end{eqnarray}
The strategy is to then seek a series solution for $f_2$, with $f_2=\sum_{n=0}^{\infty}K_nz^n$.
Then from equation (\ref{aRabia}) one obtains
\begin{eqnarray}
f_1=\sum_{n=0}^{\infty}\frac{\alpha
K_n+(1-\lambda)(n+1)gK_{n+1}}{n-x}y^n.
\end{eqnarray}
The three-term recurrence relation
\begin{eqnarray}
a_nK_{n+1}=b_nK_n+c_nK_{n-1},
\end{eqnarray}
follows from equation (\ref{aRabib}), with
\begin{eqnarray}
a_n&=&2g\sqrt{\lambda}(n+1)+\frac{(1-\lambda)(n+1)}{n-x}g\alpha,\\
b_n&=&n+(4g^2\lambda-2\frac{1-\lambda}{1+\lambda}\Delta-x)-\frac{\alpha^2}{n-x}-\frac{(1-\lambda)^2g^2n}{n-1-x},\\
c_n&=&-2g\sqrt{\lambda}-\frac{(1-\lambda)g\alpha}{n-1-x}.
\end{eqnarray}
The regular parts of the energy spectrum are given by the zeros of the $G$-functions,
$G_\pm^\lambda = 0$, defined by
\begin{eqnarray}
G_+^\lambda&=&(1+\sqrt{\lambda})(\varphi_1(-z)-\varphi_2(z))-(1-\sqrt{\lambda})(\varphi_1(z)+\varphi_2(-z)),\\
G_-^\lambda&=&(1+\sqrt{\lambda})(\varphi_1(-z)+\varphi_2(z))+(1-\sqrt{\lambda})(\varphi_1(z)-\varphi_2(-z)).
\end{eqnarray}
It is clearly seen that for $\lambda=1$ the results given for the quantum Rabi model are recovered.

\subsection{Two-photon quantum Rabi model}

The two-photon quantum Rabi model is a direct generalization of the quantum Rabi model, with hamiltonian
\begin{equation}
H=\Delta
\sigma_z+\omega a^{\dagger}a+g\sigma_x(a^{\dagger 2}+a^2).
\end{equation}
This model can be used to describe a two-level atom interacting with squeezed light \cite{Gerry} and quantum dots
inserted in a QED microcavity \cite{dValle}.
Most recently, an implementation of the two-photon quantum Rabi model has been proposed using trapped ions \cite{2photon}.
The eigenspectrum of this model has been obtained and discussed via the $G$-function approach by
various authors \cite{CWHLW2012,T2012,Maciejewski,Cui,Peng2013,T2015,DXBC}.
Some isolated exact solutions have also been obtained~\cite{ZhangY,Zhang2015,Emarya,Dolyaa,Dolyab}.

The eigenstate $\left|\psi\right\rangle$ of the two-photon quantum Rabi hamiltonian $H$ is again
\begin{equation}
\left|\psi\right\rangle=\left(
\begin{array}{c}
\psi_1 \\
\psi_2
\end{array}%
\right).
\end{equation}
The Schr\"{o}dinger equation $H|\psi\rangle=E|\psi\rangle$ gives
\begin{eqnarray}
 a^{\dagger}a\psi_1+g(a^{\dagger 2}+a^2)\psi_2+\Delta \psi_1 &=&E\psi_1,\\
  a^{\dagger}a\psi_2-g(a^{\dagger 2}+a^2)\psi_1-\Delta \psi_2 &=&E\psi_2.
\end{eqnarray}
The linear combinations $\phi_1=\psi_1+\psi_2$ and $\phi_2=\psi_1-\psi_2$ then satisfy
\begin{eqnarray}
 a^{\dagger}a\phi_1+g(a^{\dagger 2}+a^2)\phi_1+\Delta \phi_2 &=&E\phi_1,\\
  a^{\dagger}a\phi_2+g(a^{\dagger 2}+a^2)\phi_2+\Delta \phi_1 &=&E\phi_2,
\end{eqnarray}
which can be written in the matrix form
\begin{equation}
H'\left(
\begin{array}{c}
\phi_1 \\
\phi_2
\end{array}%
\right)=E\left(
\begin{array}{c}
\phi_1 \\
\phi_2
\end{array}%
\right)
\end{equation}
with
\begin{equation}
H'=\left(
\begin{array}{cc}
a^{\dagger}a+g(a^{\dagger 2}+a^2)  & \Delta \\
\Delta & a^{\dagger}a-g(a^{\dagger 2}+a^2) %
\end{array}%
\right).
\end{equation}

A set of coupled second-order differential equations follows in the Bargmann-Fock space,
\begin{eqnarray}
\phantom{-} g\frac{{d^2\phi_1}}{{dz^2}}+ z\frac{{d\phi_1}}{{dz}}+(z^2-E)\phi_1+ \Delta \phi_2=0,\label{tprabia}\\
-g\frac{{d^2\phi_2}}{{dz^2}}+ z\frac{{d\phi_2}}{{dz}}-(z^2+E)\phi_2+
\Delta \phi_2=0.\label{tprabib}
\end{eqnarray}
These equations have two symmetries \cite{T2012}:
$z\rightarrow -z$ doesn't change the equations, thus $\phi_{1,2}(z)$ are either both
even, or both odd;
$z\rightarrow \mathrm{i} z$ swaps $\phi_1(z)$ and $\phi_2(z)$, leading to the relations
\begin{equation}
\phi_1(\mathrm{i} z)=C\phi_2(z), \quad \phi_2( \mathrm{i} z)=C\phi_1(z).
\end{equation}
The different symmetries thus give four possible values of $C$,
\begin{equation}
C=\pm 1, \pm \mathrm{i}.
\end{equation}

\begin{figure}[tb]
\begin{center}
\includegraphics[width=0.8\columnwidth]{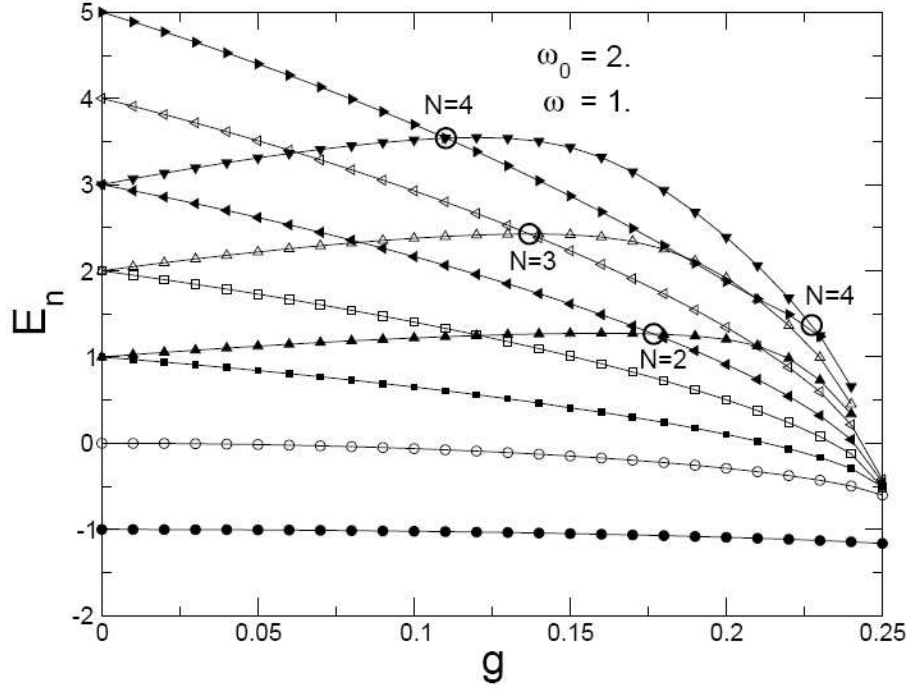}
\caption{Energy spectrum of the two-photon quantum Rabi model obtained from the $G_C(E)$-function approach
for $\Delta=\omega_0/2=1$ and $\omega=1$. The circles marked by $N=2,3,4$ correspond to exact energies
from equation (\ref{tenergy}). The corresponding values of $g$ are determined by the parameter relations
(\ref{tparaa})-(\ref{tparac}). Reproduced with permission from~\cite{T2012}.}\label{tRabien}
\end{center}
\end{figure}

Following Braak's approach \cite{Braak} for the quantum Rabi model,
the solutions $\phi_{1,2}(z)$ are first constructed.
The transformation $\phi_{1,2}(z)=e^{-\kappa z^2}\varphi_{1,2}(z)$ with $\kappa=(1-\sqrt{1-4g^2})/4g$ leads to
\begin{eqnarray}
\qquad \qquad g\frac{{d^2\varphi_1}}{{dz^2}}+
(1-4g\kappa)z\frac{{d\varphi_1}}{{dz}}-(2g\kappa+E)\varphi_1+ \Delta \varphi_2&=&0,\label{BTcepa}\\
-g\frac{{d^2\varphi_2}}{{dz^2}}+ (1+4g\kappa)z\frac{{d\varphi_2}}{{dz}}-(4\kappa z^2-2g\kappa+E)\varphi_2+ \Delta \varphi_1&=&0.\label{BTcepb}
\end{eqnarray}
Then follows the expansion for $\varphi_{1,2}(z)$,
\begin{equation}
\varphi_1(z)=\sum_{n=0}^\infty Q_n(E)z^n, \qquad \varphi_2(z)=\sum_{n=0}^\infty K_n(E)z^n.
\end{equation}
Inserting these expansions into equations (\ref{BTcepa}) and (\ref{BTcepb}) gives the iteration relations
\begin{eqnarray}
g(n+2)(n+1)Q_{n+2}+[(1-4g\kappa)n-2g\kappa-E]Q_n+\Delta K_n&=&0,\\
g(n+2)(n+1)K_{n+2}-[(1+4g\kappa)n+2g\kappa-E]Q_n &&\nonumber\\
\qquad \qquad \qquad \qquad \qquad \qquad \qquad \qquad + \, 4 \kappa K_{n-2}-\Delta Q_n&=&0.
\end{eqnarray}
It is found that the indices $n$ differ by $0, 2$, or $4$, thus only the coefficients $Q_n$ and $K_n$
with a common parity are non-zero.
A single $G$-function may be constructed for each value of the symmetry parameter $C$,
\begin{eqnarray}
 C=1, \phantom{-} \quad G_+(z,E)=\phi_2(\mathrm{i}z)-\phi_1(z)=0, \quad \quad Q_0=K_0=1,\\
 C=-1, \quad G_-(z,E)=\phi_2(\mathrm{i}z)+\phi_1(z)=0, \quad \quad Q_0=-K_0=1,\\
 C=\mathrm{i}, \phantom{-} \quad \, G_\mathrm{i}(z,E)=\mathrm{i} \phi_2(\mathrm{i} z)+\phi_1(z)=0, \quad \quad Q_1=K_1=1,\\
 C=\mathrm{-i}, \quad \, G_{-\mathrm{i}}(z,E)=\mathrm{i} \phi_2(\mathrm{i} z)-\phi_1(z)=0, \quad \quad Q_1=-K_1=1.
\end{eqnarray}
The regular part of the energy spectrum has been determined from $G_C(z,E)$, as shown in figure~\ref{tRabien}.
Note that, differently to the $G$-functions of the quantum Rabi model,
the $G_C(z,E)$ functions in fact vanish identically~\cite{Maciejewski} and thus the
problem is not well justified.
Nevertheless, the method for finding the energy spectra via the
$G_C(z,E)$ functions and the above iteration relations still works
by mixing entire functions with truncated ones~\cite{T2015}.  In
addition, the analytical approach considered here  can be used to
obtain the spectrum of  the quantum Rabi model (\ref{Rabiha1}) with
an additional nonlinear coupling $\pm\omega\sigma_za^{\dagger}a$
\cite{Maciejewskic}, since the equations for  the wave function
components in this nonlinear quantum Rabi model have a similar form
with  equations (\ref{tprabia}) and (\ref{tprabib}) with $\Delta=0$
\cite{Maciejewskic}.

For this model the exceptional eigenenergies have the form~\cite{Emarya}
\begin{equation}
\frac{E}{\omega}=-\frac{1}{2}+(N+\frac{1}{2})\sqrt{1-4\frac{g^2}{\omega^2}}, \qquad N=2,3,4,....\label{tenergy}
\end{equation}
They are valid under relations satisfied by the system parameters.
For example,
\begin{eqnarray}
N=2, \qquad \frac{g^2}{\omega^2}+\frac{\Delta^2}{24\omega^2}-\frac{1}{6}=0,\label{tparaa}\\
N=3, \qquad \frac{g^2}{\omega^2}+\frac{\Delta^2}{40\omega^2}-\frac{1}{10}=0,\label{tparab}\\
N=4, \qquad \frac{g^4}{\omega^4}-\frac{2g^2}{7\omega^2}+\frac{\Delta^4}{4480\omega^4}
-\frac{\Delta^2}{224\omega^2}+\frac{17g^2\Delta^2}{560\omega^4}+\frac{1}{70}=0.\label{tparac}
\end{eqnarray}
These exceptional energies occur at level-crossings, but do not cover every crossing
under certain parameter conditions, as seen in figure \ref{tRabien}.
The remaining energy level crossings take  the form~\cite{Dolyaa,Dolyab}
\begin{equation}
\frac{E}{\omega}=-\frac{1}{2}+N\sqrt{1-4\frac{g^2}{\omega^2}}, \qquad N=2,3,4,....
\end{equation}
For example, for $N=2$ and $N=3$,
\begin{eqnarray}
256\frac{g^2\Delta^2}{\omega^4}-(4\frac{\Delta^2}{\omega^2}-9)(1-4\frac{\Delta^2}{\omega^2})=0,
\quad \frac{1}{2}<\frac{\Delta}{\omega}<\frac{3}{2},\\
64\frac{g^2\Delta}{\omega^2}-(5\mp2\frac{\Delta}{\omega})(2\frac{\Delta}{\omega}\pm 3)(1-\pm 2\frac{\Delta}{\omega})=0,
\quad \frac{1}{2}<\frac{\Delta}{\omega}<2\pm\frac{1}{2},
\end{eqnarray}
respectively.
These exceptional energies occur for $\Delta/\omega>1/2$.

The two-photon Rabi model can also be solved with the Bogoliubov transformation.
A concise form of the $G$-functions obtained in this way \cite{CWHLW2012} is
\begin{equation}
G_{e,o}^{\pm}=\sum_{n}f_n[1\mp\Delta\frac{u^2+\nu^2}{n-x}]L_n^{e,o}=0,
\end{equation}
where   $x=\nu^2+E(u^2+\nu^2)$ and
\begin{eqnarray}
L_{n=2k}^e=\frac{(2k)!(u\nu)^k}{2^k}\sum_{j=0}^k\frac{(-\frac{\nu^2}{u^2})^j}{j!(k-j)!},\\
L_{n=2k+1}^o=\frac{(2k+1)!\nu(u\nu)^k}{2^k}\sum_{j=0}^k\frac{2^{2j}j!(-\frac{\nu^2}{u^2})^j}{(2j+1)!(k-j)!}.
\end{eqnarray}
Here $u=\sqrt{(\beta+1)/2}$, $\nu=\sqrt{(\beta-1)/2}$ and $\beta=1/\sqrt{1-4g^2}$.
The coefficients $f_n$ satisfy the three-term relation
\begin{equation}
(m+2)(m+1)f_{m+2}=\frac{\Omega(m)}{u\nu+g(u^2+\nu^2)})f_{m}-f_{m-2},
\end{equation}
with
\begin{equation}
\Omega(m)=(u^2+\nu^2)m+\nu^2+2gu\nu(2m+1)-E-\frac{\Delta}{\frac{m-\nu^2}{u^2+\nu^2}-E}.
\end{equation}

The $G$-function for this model has been more recently derived in a concise and compact way
by using extended squeezed states for each Bargmann-Fock index \cite{DXBC}.
The anisotropic version of the two-photon quantum Rabi model, in which the rotating and counter-rotating
terms enter with different coupling constants, has also been solved by
employing a variation of the Braak method based on Bogoliubov rotation of the underlying $su(1, 1)$ Lie algebra \cite{Cui}.

\section{Experimental realisation}

The quantum Rabi model hamiltonian (\ref{Rabiha1}) can be realised in many physical systems,
including  cavity systems \cite{RBH2001,Walther2006},
superconducting circuits \cite{Makhlin, Blais,Buluta},
trapped ions \cite{Liebfried},
quantum dots~\cite{Englund}
and hybrid quantum systems \cite{Wallquist,Aspelmeyer,IS}.
In this Section we give a brief outline of some key experimental realisations of the
quantum Rabi model in such distinctly different physical systems.

\subsection{Single atom in a cavity}

\begin{figure}[tb]
\begin{center}
\includegraphics[width=1.0\columnwidth]{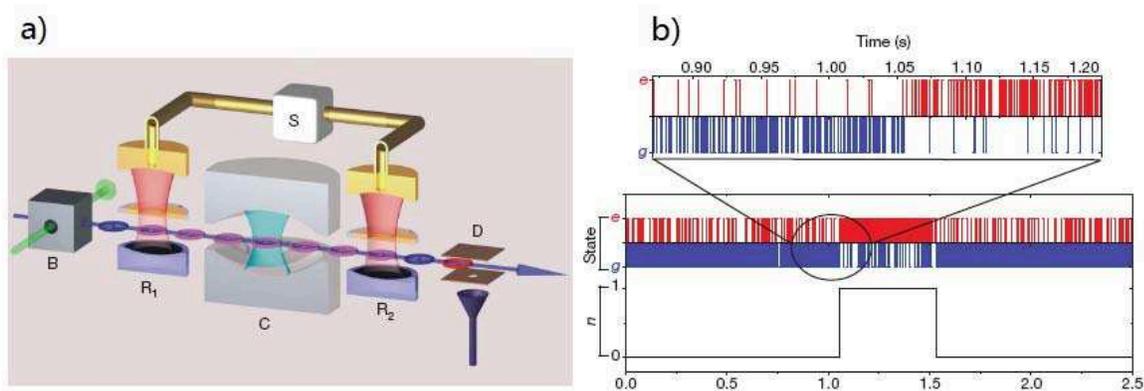}
\caption{a) The experimental setup for a single atom in a cavity. In box B, the circular Rydberg atoms are
prepared in the circular state $|g\rangle$. The atoms cross the cavity $C$ which is  sandwiched by two
Ramsey cavities $R_1$ and $R_2$ fed by the classical microwave source $S$.
They are detected in the state selective field ionization detector $D$.
b) Quantum jump of the photons. Here red and blue bars are  the raw signal,
a sequence of atoms detected in $|e\rangle$ or $|g\rangle$, respectively.
A sudden change occurs in the statistics of the detection events, revealing the quantum
jump of the photon from $|0\rangle$ to $|1\rangle$. Reproduced with permission from~\cite{Gleyzes}.}\label{experimenta}
\end{center}
\end{figure}

The quantum Rabi model can be realised  with a single atom in cavity.
For example, in the  microwave cavity quantum electrodynamics
experiments \cite{Gleyzes,Guerlin}, the dispersive interaction of single atoms
with the trapped intra-cavity-photons in the strong coupling regime has been realised.
The atom is in the circular Rydberg states with two high principle quantum numbers $n=50$ and $n=51$.
The two circular levels are denoted by  $|g\rangle$ and  $|e\rangle$  with energies $E_{g,e}$.
The cavity is constructed using superconducting niobium mirrors  marked by $C$ in figure \ref{experimenta}a,
and  sustains  a single mode radiation field of  the frequency $\omega$.

In the dipole approximation, the interaction of the atom and cavity can be described by the hamiltonian
\begin{equation}
H=\hbar \Delta(|e\rangle\langle e|-|g\rangle\langle g|)+\hbar \omega a^{\dagger}a
+\hbar g(a^\dagger+a)(|g\rangle\langle e|+|e\rangle\langle g|),
\end{equation}
with $\Delta=(E_e-E_g)/2\hbar$ and $g=d\sqrt{(\hbar \omega)/2\epsilon_0V}/\hbar$.
Here $d$ is  the dipole matrix element for the transition between the two levels and $V$ is  the volume of the cavity.
Using the Pauli spin operators to represent the atomic operators,
\begin{equation}
\sigma_z=|e\rangle\langle e|-|g\rangle\langle g|,\quad \sigma_x=|e\rangle\langle g|+|g\rangle\langle e|,
\end{equation}
gives
\begin{equation}
H=\hbar \Delta\sigma_z+\hbar \omega a^{\dagger}a+\hbar g(a^\dagger+a)\sigma_x.
\end{equation}

In such cavity quantum electrodynamics experiments, the parameter values are
$\omega/2\pi=51.1$GHz and $g/2\pi=51$kHz, with the ratio $g/\omega$ given by
\begin{equation}
{g}/{\omega}\approx 10^{-6}.
\end{equation}
For this small value of $g/\omega$, the interaction between the atom and field can be described by the JC model.
Much larger coupling strengths of relevance to the full Rabi model can be achieved in experiments
across other platforms.
In this experiment, the non-destructive measurements on single photons are performed via atoms.
Quantum jumps in the photon number are observed (see figure \ref{experimenta}b).

\subsection{Superconductor circuits}

\begin{figure}[tb]
\begin{center}
\includegraphics[width=1.0\columnwidth]{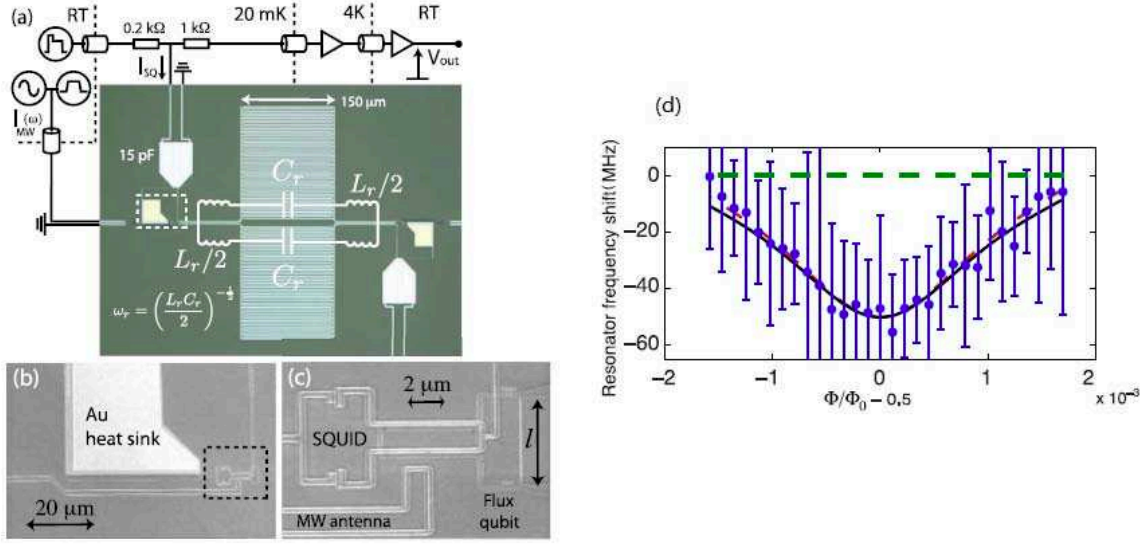}
\caption{(a) Schematic diagram of the quantum circuit. The LC resonator is made of two capacitors,
each containing $50$ figure of $150\mu$m and $1.5\mu$m width,
and linked by two long superconducting wires. There are two SQUIDs next to the resonator.
(b) Scanning electron micrograph (SEM) picture of the SQUID circuit.
(c) SEM picture of the qubit coupling to the resonator by the coupling wire of length $l$.
(d) Bloch-Siegert Shift. The blue dots are the measurement results
and the dashed green lines are the prediction from the JC model.
The solid black line is the fit from the quantum Rabi model
and the dashed red line is the approximate result in Ref.~\cite{Mooij}.
Reproduced with permission from~\cite{Mooij}.}\label{circuit}
\end{center}
\end{figure}

The quantum Rabi model is also realised in superconductor  circuits.
In particular, in recent experiments where a flux qubit is coupled to an $LC$ resonator,
the ultrastrong-coupling regime has been achieved~\cite{Mooij}.
The flux qubit consists of four Josephson junctions interrupting a superconducting loop (see figure \ref{circuit}c),
which is threaded by an external flux bias $\Phi$.
In certain conditions, the qubit potential landscape is of the form of a double-well potential,
where the two minima correspond to states with clockwise and anticlockwise persistent currents $\pm I_p$.
The qubit thus behaves effectively as a two-level system in the basis of the two persistent states,
with effective hamiltonian
\begin{equation}
H_q=-\hbar\epsilon \sigma_z- \hbar \Delta\sigma_x,
\end{equation}
where $\epsilon=4I_p(\Phi-\Phi_0/2)/\hbar$ with the flux quantum  $\Phi_0=h/2e$
and $2\Delta$ is the tunneling coupling between the two persistent states.
The resonator is made of two capacitors (see figure \ref{circuit}a)
and can be described as the harmonic oscillator
\begin{equation}
H_r=\hbar\omega_r a^{\dagger}a,
\end{equation}
where $\omega_r=1/\sqrt{L_rC_r/2}$ is the resonance frequency.

The qubit is galvanically attached to the resonator with a coupling wire (see figure \ref{circuit}c).
The interaction between the qubit and the resonator is described by
\begin{equation}
H_{int}=\hbar g(a^\dagger+a)\sigma_z,
\end{equation}
where $g=I_pI_{rms}L_K/\hbar$ with the kinetic inductance $L_K=0.14\hbar R_n/k_BT_c$
and the zero-point current fluctuation $I_{rms}=\sqrt{\hbar\omega_r/2L_r}$.
For typical parameter values in  this experiment, the ratio $g/\omega_r$ is
\begin{equation}
{g}/{\omega_r}=I_pI_{rms}L_K\sqrt{L_rC_r/2}/\hbar\approx 0.1,
\end{equation}
which belongs to the ultrastrong-coupling regime.
In this experiment a $50$MHZ Bloch-Siegert shift has been measured (see figure \ref{circuit}d).

\begin{figure}[tb]
\begin{center}
\includegraphics[width=1.0\columnwidth]{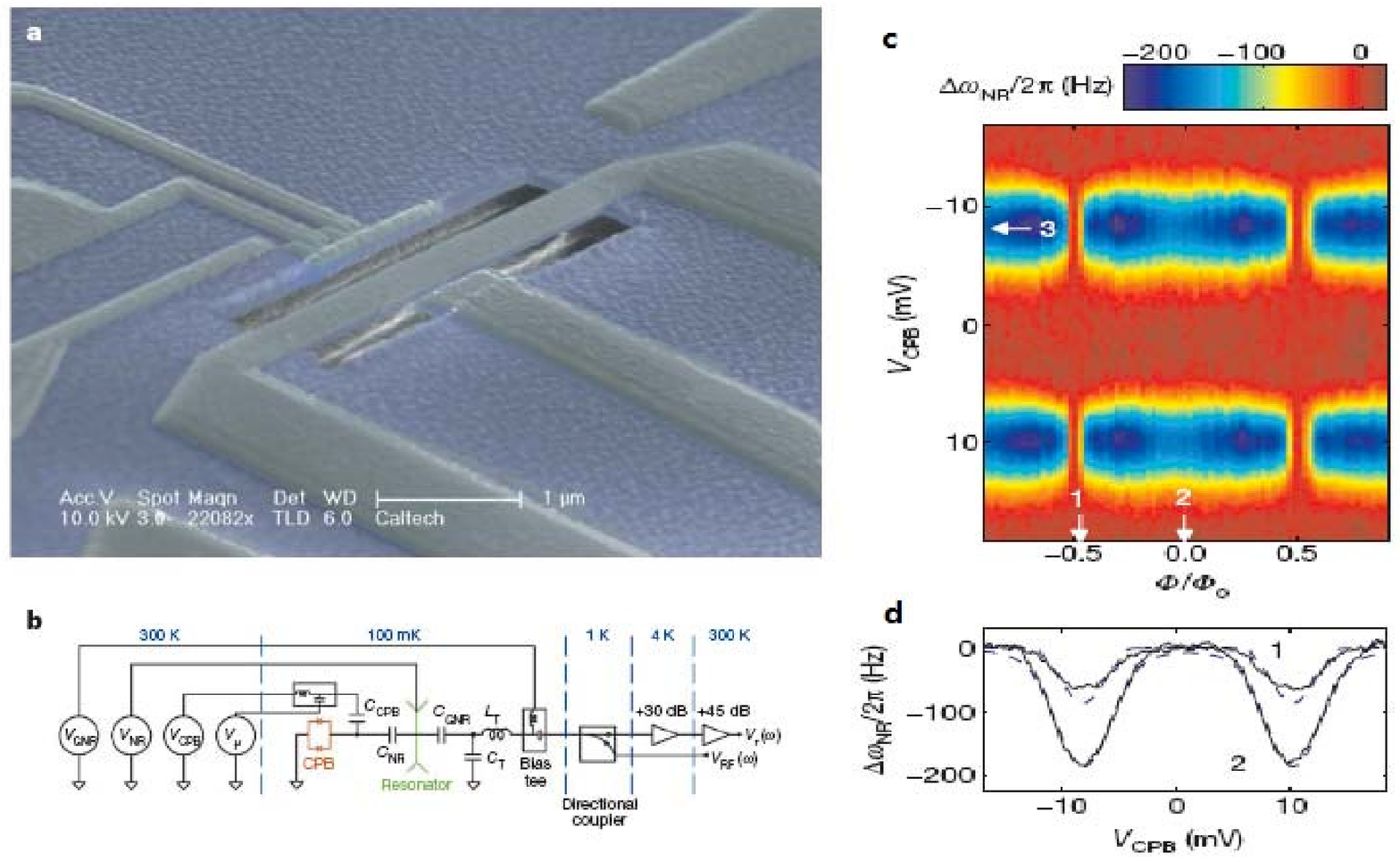}
\caption{(a) Colourised scanning electron micrograph of the devices.  The nanoresonator is
formed from low-stress silicon nitride with a thin coating of aluminium for applying $V_\mathrm{NR}$.
The CPB is positioned at a distance $\sim 300$nm from the nanoresonator.
(b) Circuit schematic for measuring the dispersive shift $\Delta \omega_\mathrm{NR}/2\pi$
of the nanomechanical frequency using radiofrequency reflectometry.
(c) Measured nanoresonator frequency shift as a function of CPB parameters $V_\mathrm{CPB}$ and
$\Phi/\Phi_0$ for $V_\mathrm{NR}=7$V.
(d) Comparison between data (solid black lines) and model (dashed blue lines)
of selected traces of $\Delta \omega_\mathrm{NR}/2\pi$ versus $V_\mathrm{CPB}$.
Here the $\Phi$ bias is near minimum $E_J$ (labelled by 1) and maximum $E_J$
(labelled by 2) in figure c.  Reproduced with permission from~\cite{LaHaye}.}\label{Nanore}
\end{center}
\end{figure}

\subsection{Hybrid quantum systems}

The quantum Rabi model can also be realised in the hybrid systems~\cite{Wallquist,Aspelmeyer}
which integrate different physical systems.
Here we discuss hybrid mechanical systems in which a mechanical oscillator is coupled to superconducting qubits,
as realised in an experiment~\cite{LaHaye}.
The nanomechanical resonator is  the fundamental in-plane
flexural mode of a suspended silicon nitride nanostructure (see figure \ref{Nanore}a),
which can be described by the harmonic oscillator
\begin{equation}
H_\mathrm{NR}= \hbar \, \omega_\mathrm{NR} \, a^{\dagger}a.
\end{equation}
In experimental conditions, $\omega_\mathrm{NR}/2\pi=58$MHz
and the effective mass of the resonator is $M\approx 4\times 10^{-6}$kg.
The superconducting qubit is a Cooper Pair Box (CPB) coupled to the nanoresonator though capacitance $C_\mathrm{NR}$.
The dynamics of the CPB can be restricted to the two energetically lowest charge states $|n\rangle$ and $|n+1\rangle$,
and can be well described by the simple spin-$\frac{1}{2}$ hamiltonian
\begin{equation}
H_\mathrm{CPB}= \frac{E_{el}}{2} \sigma_z-\frac{E_J}{2}\sigma_x.
\end{equation}
In the first term of $H_\mathrm{CPB}$, $E_{el}=8E_C(n_\mathrm{CPB}+n_\mathrm{NR}-n-1/2)$ is the
electrostatic energy difference between the charge states
and $E_C=e^2/2(C_\mathrm{NR}+C_\mathrm{CPB}+2C_J)$ is the charge energy,
where $C_J$ is the capacitance of the
Josephson junction, $C_\mathrm{CPB}$ is the capacitance between the CPB
island and a nearby gate electrode, and  $n_\mathrm{CPB}=C_\mathrm{CPB} V_\mathrm{CPB}$
and $n_\mathrm{NR}=C_\mathrm{NR} V_\mathrm{NR}$
are the polarization charges applied by the gate electrode and the nanoresonator, which
are held at potentials $V_\mathrm{CPB}$ and $V_\mathrm{NR}$, respectively.
In the second term of $H_\mathrm{CPB}$, $E_J=E_{J0}|\cos(\pi \Phi/\Phi_0)|$
is the Josephson energy of the junctions, where $\Phi$ the externally applied magnetic flux.

The small displacement $x$ of the nanoresonator can result in  linear modulation
of the capacitance between the nanoresonator and CPB, with
$C_\mathrm{NR}=C_\mathrm{NR}(0)+(\partial C_\mathrm{NR}/\partial x)x$.
The resulting interaction hamiltonian takes the form
\begin{equation}
H_{int}=\hbar g(a^\dagger+a)\sigma_z,
\end{equation}
where
\begin{equation}
g=\frac{4n_{NR}E_c}{\hbar}\frac{1}{C_{NR}}\frac{\partial C_{NR}}{\partial x}x_{zp},
\end{equation}
and  $x_{zp}=\sqrt{\hbar/2M\omega_{NR}}$.
In this experiment, a dispersive shift of the nanomechanical frequency  has been observed (see figures \ref{Nanore}c and \ref{Nanore}d).

\subsection{Artificial atoms in the ultrastrong and deep strong coupling regimes}

The superconducting flux qubits are regarded as artificial atoms.
The architecture of circuit QED, made up of Josephson junctions, superconducting waveguides, inductors and capacitors,
is used in the most recent experiments~\cite{Forn-Diaz,Yoshihara} to simulate atoms in the ultrastrong and deep strong
coupling regimes.
In different setups,
Forn-Diaz {\sl et al.}~\cite{Forn-Diaz} couple the flux qubit to a transmission line and
Yoshihara {\sl et al.}~\cite{Yoshihara} couple the flux qubit to a microwave resonator.
Also using circuit QED, a multi-cell photonic crystal with a bandgap structure has been realised
by alternating segments of waveguide with varying impedance~\cite{LH2016}.
Replacing the transmon qubit with a flux qubit promises to push the photonic crystal into the
ultrastrong and deep strong coupling regimes.
The key point, as touched upon in this Section, is that such systems allow the simulation of the
quantum Rabi model, including the asymmetric version, in all coupling regimes.
Quantum simulators of this kind are thus exciting platforms for future discoveries in fundamental models of
light-matter interaction.

\section{Concluding remarks}

In this article we have reviewed recent progress on the quantum Rabi model and related generalised models --
the asymmetric quantum Rabi model, the  anisotropic quantum Rabi model and the two-photon quantum Rabi model.
Our emphasis has been on the different analytic methods used to solve these models
and the analytic solutions obtained for their full eigenspectrum.
These solutions have been obtained over the past five years following Braak's analytic solution \cite{Braak} of the quantum Rabi model.
We have also discussed some results for the dynamics of the quantum Rabi model and presented an overview of
experimental realisations in different experimental settings.

The analytic solutions apply in the full parameter space of the models,
in parameter regimes which have traditionally challenged various numerical approaches.
A number of immediate questions arise regarding how these analytic solutions can be applied.
We list some of these questions here.
For example, how can the analytic solutions be applied to the calculation of the dynamics, to the calculation of
Berry and other geometric phases,
and to the calculation of other quantities of interest like fidelity and multipartite entanglement?
What is the connection, if any, between the analytic solution and
previous work on `solving' the quantum Rabi model in terms of spheroidal wave functions~\cite{Reika}?
The JC model has been solved for a general parafermion replacing the usual qubit~\cite{para}.
Can the analytic solution of the quantum Rabi model be extended to a parafermion?
The physics of an $N$-state atom interacting with a light field is rich~\cite{Albert2012}.
Some initial results for the analytic solution of the quantum Rabi model extended to an $N$-state atom
have been obtained~\cite{ZhangY2014}.
Much work also remains to fully develop the analytic solution of the Rabi model for multi-qubits,
with some specific analytic solutions already given~\cite{Peng2015,Peng2014,Chilingaryan2013}.
The bipartite and multipartite entanglement dynamics of multi-qubit systems is of
considerable interest (see, e.g.,~\cite{ZCHC,MLZ2016b}).

The overarching challenge is to uncover new phenomena to test the experimental and
quantum simulation advances which are now capable of probing the ultrastrong and deep strong coupling regimes.
One promising theoretical development which may have interesting experimental consequences on the quantum information
side is the discovery of robust GHZ-like dark-states built from isolated exact solutions~\cite{Peng2015,Peng2014}.
As stated elsewhere~\cite{Rabipreface}, the future of the quantum Rabi model and related models is very bright.
We look forward to further developments.

\ack

It is a pleasure to thank a number of colleagues who have influenced our understanding of various aspects of the quantum Rabi and related models, both through their publications and in discussions.
We mention in particular, Daniel Braak, Qing-Hu Chen, Andrzej Maciejewski, Maria Przybylska, Enrique Solano, Yun-Bo Zhang, and Hong-Gang Luo.
This work was supported by the National Basic Research Program of China (Grant No. 2012CB821305),
the National Natural Science Foundation of China (Grants No. 11374375, 11574405, 11375059, 11565011, 11465008 and 11174375)
and Australian Research Council Discovery Project DP170104934.
This work of MTB was also been partially supported by QianRen Program of China.

\appendix

\section{Confluent Heun equation and confluent Heun function}

Here we give further details of the derivation of the analytic solutions (\ref{af1a})-(\ref{af2b}) obtained in terms of
confluent Heun functions.
In the notation adopted in Maple, the confluent Heun equation is defined by
\begin{equation}
\frac{d^2 y}{dz^2} + \left( \alpha + \frac{\beta+1}{z} + \frac{\gamma+1}{z-1} \right) \frac{dy}{dz} +
\left( \frac{\mu}{z} + \frac{\nu}{z-1} \right) y = 0 .
\nonumber
\end{equation}
The confluent Heun equation follows from the Heun equation when one of the three regular finite singularities
becomes coincident (confluent) with the singularity at $\infty$ (see, e.g., \cite{Ronveaux,Slavyanov}).
The solution is an infinite series
\begin{equation}
\mathrm{HC}(\alpha,\beta,\gamma,\delta,\eta,z) = \sum_{n=0}^\infty v_n(\alpha,\beta,\gamma,\delta,\eta) z^n,
\end{equation}
which is regular around the singular point $z=0$.
The coefficients $v_n$ obey the three-term recurrence relation
\begin{equation}
A_n v_n = B_n v_{n-1} + C_n v_{n-2},
\end{equation}
with $v_{-1}=0$, $v_0=1$ and
\begin{eqnarray}
A_n&=&1+\beta/n,\\
B_n&=&1+(\beta+\gamma-\alpha-1)/n+[\eta-\beta/2+(\gamma-\alpha)(\beta-1)/2]/n^2,\\
C_n&=&[\delta+\alpha(\beta+\gamma)/2+\alpha(n-1)]/n^2.
\end{eqnarray}

With the above results in view,
we first consider the solution for $\phi_1(z)$ in (\ref{Bseqa}) which admits two types of solutions, namely
\begin{eqnarray}
&\textrm{Type-I:}~~~& \phi_{1}(z)=e^{-gz} f_1(x_1),\\
&\textrm{Type-II:}~~& \phi_{1}(z)=e^{gz} f_2(x_2),
\end{eqnarray}
where $x_1=(g-z)/2g$ and $x_2=(g+z)/2g$.
Substitution of $\phi_1(z)=e^{-gz} f_1(x_1)$ into equation (\ref{Bseqa}) gives a confluent Heun equation for $f_1(x_1)$,
\begin{equation}
\frac{d^2f_1}{dx_1^2}+\left(\alpha_1+\frac{\beta_1+1}{x_1} +\frac{\gamma_1+1}{x_1-1}\right)
\frac{df_1}{dx_1}+\frac{\mu_1 x_1+\nu_1}{x_1(x_1-1)}f_1=0,
\label{heun1}
\end{equation}
where $\mu_1=\delta_1+\alpha_1(\beta_1+\gamma_1+2)/2$ and $\nu_1=\eta_1+\beta_1/2+(\gamma_1-\alpha_1)(\beta_1+1)/2$.
The parameters $\alpha_1, \beta_1,\gamma_1,\delta_1$ are as given in the main body of the text.

If  $-\beta_1=(E+g^2+1)$ is not zero and a negative integer, the confluent Heun equation
has two linearly independent local Frobenius solutions around $x_1=0$~\cite{Ronveaux,Slavyanov}.
These are
\begin{eqnarray}
f_1^1(x_1)&=&\textrm{HC}(\alpha_1,\beta_1,\gamma_1,\delta_1,\eta_1,x_1),\\
f_1^2(x_1)&=&x_1^{-\beta_1}\textrm{HC}(\alpha_1,-\beta_1,\gamma_1,\delta_1,\eta_1,x_1),
\end{eqnarray}
where the confluent Heun function HC is defined in (\ref{HCdef}).
If both $\phi_1^1(x_1)$ and $\phi_1^2(x_1)$ are regarded as physically acceptable solutions,
they must be entire functions over the whole complex plane.
$f_1^1(x_1)$ satisfies such a requirement at least formally, since $f_1^1(x_1)$ is a series in $x_1=(g-z)/2g$.
$\phi_1^2(x_1)$ may represent a physical solution only when $-\beta_1$ is  a positive integer because of the
presence of the term $x_1^{-\beta_1}$.
Therefore, in a general situation, $f_1^1(x_1)$ can be used to construct the solution
\begin{equation}
f_1(z)=C_1\, e^{-gz} \,
\textrm{HC}\left(\alpha_1,\beta_1,\gamma_1,\delta_1,\eta_1,\frac{g-z}{2g}\right),
\label{f1a}
\end{equation}
for $\phi_1(z)$, where $C_1$ is a constant to be determined.

On the other hand, substituting $\phi_1(z)=e^{gz} f_2(x_2)$ into equation (\ref{Bseqa}),
gives a similar confluent Heun equation for $f_2(x_2)$,
\begin{equation}
\frac{d^2f_2}{dx_2^2}+\left(
\alpha_2+\frac{\beta_2+1}{x_2}+\frac{\gamma_2+1}{x_2-1}\right) \frac{df_2}{dx_2} +\frac{\mu_2 x_1+\nu_2}{x_2(x_2-1)}f_2=0,
\end{equation}
where $\mu_2=\delta_2+\alpha_2(\beta_2+\gamma_2+2)/2$ and $\nu_2=\eta_2+\beta_2/2+(\gamma_2-\alpha_2)(\beta_2+1)/2$.
The parameters $\alpha_2, \beta_2,\gamma_2,\delta_2$ are given in the main body of the text.
In a similar way, the solution for $f_2(z)$ is of the form
\begin{equation}
f_2(z)=C_2 \,
\textrm{HC} \, \left(\alpha_2,\beta_2,\gamma_2,\delta_2,\eta_2,\frac{g-z}{2g}\right),
\label{f1b}
\end{equation}
where $C_2$ is also a constant.
Therefore, we obtain two types of solution for $\phi(z)$ in terms of the confluent Heun functions,
\begin{eqnarray}
\phi_1^1(z)&=&C_1 \, e^{-g z} \, \textrm{HC}\left(\alpha_1,\beta_1,\gamma_1,\delta_1,\eta_1,\frac{g-z}{2g}\right),\label{apaf1a}\\
\phi_1^2(z)&=&C_2 \, e^{g z}\, \textrm{HC}\left(\alpha_2,\beta_2,\gamma_2,\delta_2,\eta_2,\frac{g+z}{2g}\right).\label{apaf1b}
\end{eqnarray}

Likewise $\phi_2(z)$ admits two types of solution
\begin{eqnarray}
&\textrm{Type-I:}~~~& \phi_{2}(z)=e^{-gz} \chi_1(x_1),\\
&\textrm{Type-II:}~~& \phi_{2}(z)=e^{gz} \chi_2(x_2),
\end{eqnarray}
with $\chi_{1,2}$ satisfying the confluent Heun equation.
It follows that the two types of solution for $\phi_2(z)$ are given in terms of confluent Heun functions
\begin{eqnarray}
\phi_2^1(z)&=&D_1\, e^{-g z} \, \textrm{HC}\left(\alpha_2,\beta_2,\gamma_2,\delta_2,\eta_2,\frac{g-z}{2g}\right),\label{apaf2a}\\
\phi_2^2(z)&=&D_2 \, e^{g z} \, \textrm{HC}\left(\alpha_1,\beta_1,\gamma_1,\delta_1,\eta_1,\frac{g+z}{2g}\right),\label{apaf2b}
\end{eqnarray}
for constants $D_{1,2}$.

Two sets of solutions -- (\ref{apaf1a}) and (\ref{apaf2a}), (\ref{apaf1b}) and (\ref{apaf2b}) -- have thus been constructed for
the coupled equations (\ref{Bceqa}) and (\ref{Bceqb}).
The relation between the constants $C_1$ ($C_2$) and $D_1$ ($D_2$) may be determined by the fact that
they must satisfy either equation (\ref{Bceqa}) or (\ref{Bceqb}).
Equation (\ref{Bceqb}) is used to obtain the relation $D_1/C_1=\Delta/(E+g^2)$ due to the presence of the term $z-g$.
Similarly $C_2/D_2=\Delta/(E+g^2)$ follows from equation (\ref{Bceqa}).
The constants $C_1$ and $D_2$ may be further determined by the normalisation of the wavefunctions.
We thus finally obtain the two types of analytic solutions given in equations (\ref{af1a}), (\ref{af2a}) and (\ref{af1b}), (\ref{af2b}).
For brevity we have chosen $C_1=D_2=1$.

\end{document}